\documentclass[twocolumn]{aastex631}

\usepackage{color,soul}

\newcommand{\solar}{$_{\odot}$}
\newcommand{\kms}{\,km s$^{-1}$}

\defcitealias{paper1}{Paper~I}

\shorttitle{The stellar remnant of SN~1181}
\shortauthors{Lykou et al.}
\graphicspath{{./}{figures/}}

\begin{document}

\title{A new study on a type Iax stellar remnant and its probable association with SN 1181}

\author[0000-0002-6394-8013]{Foteini Lykou}
\affiliation{Konkoly Observatory, Research Centre for Astronomy and Earth Sciences, Konkoly-Thege Mikl\'os \'ut 15-17, 1121 Budapest, Hungary}
\affiliation{Laboratory for Space Research, University of Hong Kong, 405B Cyberport 4, 100 Cyberport Road, Cyberport, Hong Kong}

\author[0000-0002-2062-0173]{Quentin A. Parker}
\correspondingauthor{Quentin A. Parker}, \email{quentinp@hku.hk}
\affiliation{Laboratory for Space Research, University of Hong Kong, 405B Cyberport 4, 100 Cyberport Road, Cyberport, Hong Kong}
\affiliation{Department of Physics, The University of Hong Kong, Chong Yuet Ming Physics Building, Pokfulam Road, Hong Kong}

\author[0000-0003-0869-4847]{Andreas Ritter}
\affiliation{Laboratory for Space Research, University of Hong Kong, 405B Cyberport 4, 100 Cyberport Road, Cyberport, Hong Kong}
\affiliation{Department of Physics, The University of Hong Kong, Chong Yuet Ming Physics Building, Pokfulam Road, Hong Kong}

\author[0000-0002-3171-5469]{Albert A. Zijlstra}
\affiliation{Jodrell Bank Centre for Astrophysics, The University of Manchester, Alan Turing Building, Oxford Road, M13 9PL, Manchester, U.K.}
\affiliation{Laboratory for Space Research, University of Hong Kong, 405B Cyberport 4, 100 Cyberport Road, Cyberport, Hong Kong}

\author[0000-0001-5094-8017]{D. John Hillier}
\affiliation{Department of Physics and Astronomy \& Pittsburgh Particle Physics, Astrophysics, and Cosmology Center (PITT PACC), University of Pittsburgh, 3941 O’Hara Street, Pittsburgh, PA 15260, USA}

\author[0000-0002-7759-106X]{Mart\'in A. Guerrero}
\affiliation{Instituto de Astrof\'isica de Andaluc\'ia (IAA-CSIC), Glorieta de la Astronom\'ia S/N, 18008 Granada, Spain}

\author[0000-0003-2385-0967]{Pascal Le D\^u}
\affiliation{Kermerrien Observatory, F-29840 Porspoder, France}

\begin{abstract}
We report observations and modeling of the stellar remnant and presumed double-degenerate merger of Type~Iax supernova Pa30, which is the probable remnant of SN~1181~AD. It is the only known bound stellar SN remnant and the only star with Wolf-Rayet features that is neither a planetary nebula central star nor a massive Pop I progenitor. We model the unique emission-line spectrum with broad, strong O~{\sc vi} and O~{\sc viii} lines as a fast stellar wind and shocked, hot gas. Non-LTE wind modeling indicates a mass-loss rate of $\sim 10^{-6}\,\rm M_\odot\,yr^{-1}$ and a terminal velocity of  $ \sim$15,000~km\,s$^{-1}$, consistent with earlier results. O~{\sc viii} lines indicate shocked gas temperatures of  $T \simeq 4\,$MK. We derive a magnetic field upper limit of $B<2.5\,$MG, below earlier suggestions. The luminosity indicates a remnant mass of 1.0--1.65\,\rm M$_\odot$ with ejecta mass $0.15\pm0.05\,\rm M_\odot$. Archival photometry suggests the stellar remnant has dimmed by $\sim$0.5 magnitudes over 100 years. A low Ne/O$\,<0.15$ argues against a O-Ne white dwarf in the merger. A cold dust shell is  only the second detection of dust in a SN Iax and the first of cold dust. Our ejecta mass and kinetic energy estimates of the remnant are consistent with Type Iax extragalactic sources. 

\end{abstract}


\section{Introduction} \label{sec:intro} 

A new class of thermonuclear supernovae (SNe), produced in low-mass (i.e., $M_* \leq 8$~M\solar) binary systems, has been identified in the last twenty years: the Type~Iax \citep[e.g.,][]{2013ApJ...767...57F}, also known as SN 2002cx-like. These are the least-understood supernovae. Two scenarios are proposed for their formation. In the single-degenerate scenario \citep[e.g.,][]{jordan2012,2015MNRAS.450.3045K}, a carbon-oxygen (CO) white dwarf (WD) that accreted material from a helium donor fails to detonate. In the double-degenerate scenario \citep[e.g.,][]{kashyap2018}, the SN is caused by deflagration of the accretion disk that forms after the merger of a CO WD with an oxygen-neon (ONe) WD companion. Type~Iax SNe are sub-luminous (as faint as $M=-13$ to $-14$ mag) with expansion velocities  from $2000$ to $9000$\kms, and may leave a stellar remnant behind. The currently known Type Iax SNe population\footnote{See also \url{https://www.wiserep.org/}, where 44 objects are currently listed.} is only a few tens \citep{2013ApJ...767...57F, 2017hsn..book..375J}, so the full range of observed properties is weakly constrained. All are extra-galactic save two in our own Galaxy. One is SN Sgr A East near the Galactic Center as suggested by \citet{2021ApJ...908...31Z}. The other\footnote{Originally assigned as Type Iax by \citet{oskinova2020}. } is associated with the nebula Pa\,30, identified as the most probable remnant of the historical SN~1181~AD \citep[][hereafter Paper I]{paper1}. 

Earlier studies show that the nebula -- Pa\,30 in \citet{kron2014} and IRAS\,00500+6713 in \citet{gvaramadze2019} -- emits strongly in X-rays \citep{oskinova2020}. The nebula appears circular, approximately 3\arcmin\ - 4\arcmin\ across depending on the wavelength of the imagery. Faint, diffuse [O {\sc iii}] nebular emission is seen via deep narrow-band imaging \citep{kron2014, oskinova2020, paper1}. An [S {\sc ii}] shell is expanding at $1000\pm100$ km~s$^{-1}$  \citepalias{paper1}. The stellar remnant, reported by \citet{gvaramadze2019} as J005311, is associated with the infrared source 2MASS J00531123+6730023.  It is located at the center of Pa\,30 and has been argued to be the result of a double-degenerate CO\,+\,ONe WD merger \citep{gvaramadze2019,oskinova2020}. 
 
In this paper we analyze the stellar wind and nebula in order to constrain the progenitor system. In the following Section, we describe our own observations as well as the archival data used in this work. Stellar parameters are derived through a comparison of the stellar spectra against NLTE and photoionization models (Sect.~\ref{sec:CS}). In Section~\ref{sec:varphot} we explore the remnant's potential variability, followed by a new analysis on the nebular properties in Sect.~\ref{sec:mass}. In the penultimate Section we discuss the implication of our results on the progenitor and its relation to other Type Iax sources, while our conclusions can be found in the last section.


\begin{figure}[tbp]
	\centering
		\includegraphics[width=\columnwidth]{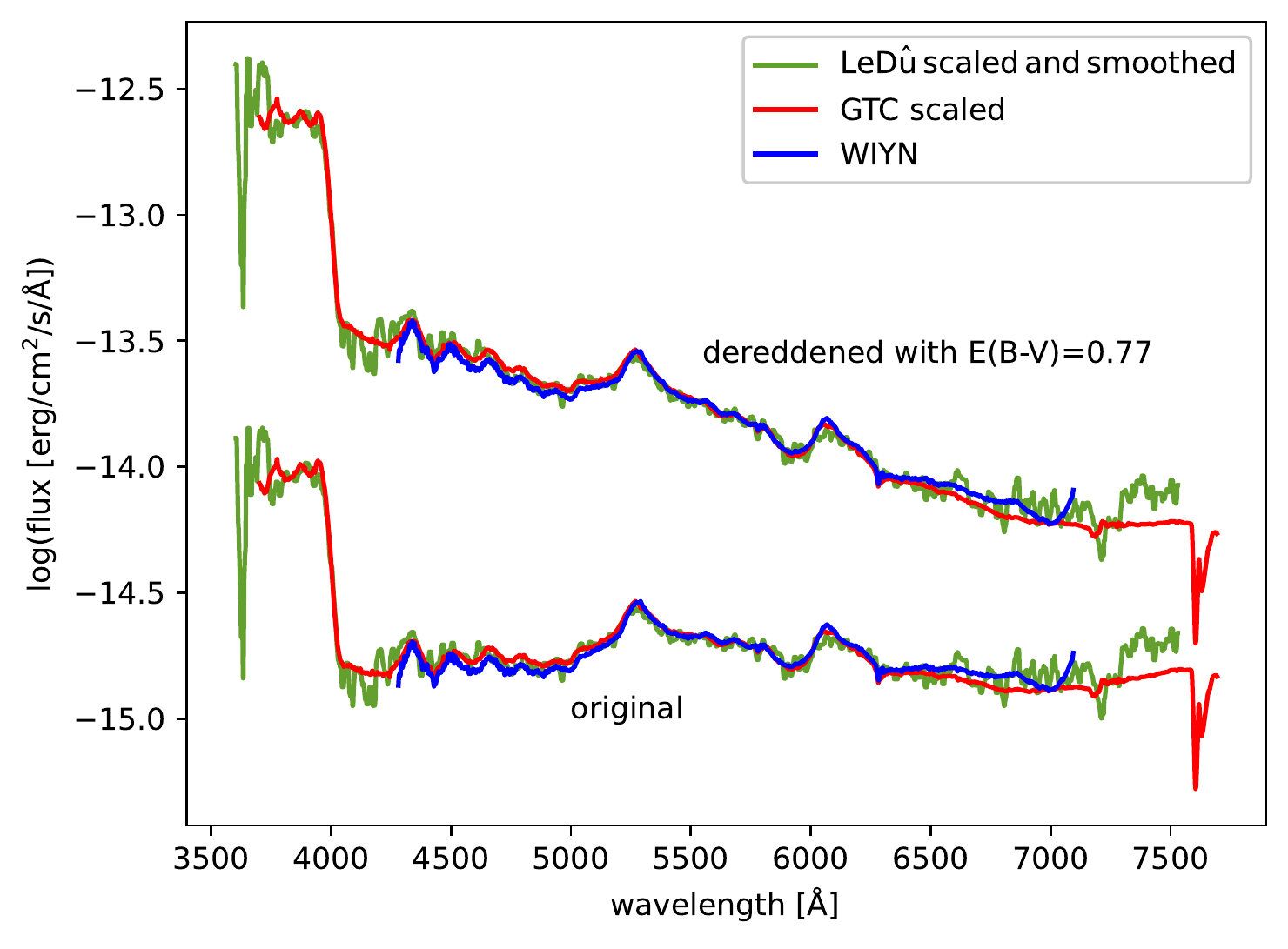}
	\caption{Stellar spectra from SparsePak/WIYN, OSIRIS/GTC, and a small amateur telescope (P.~Le D\^u) smoothed and scaled to the same flux level. The dereddened spectra are also shown for comparison \citepalias[using interstellar extinction from ][]{paper1}. The 
	major discontinuity bluewards of 4000\,\AA\ is seen in the OSIRIS and amateur spectra but the SparsePak spectrum did not go below 4200\,\AA. The OSIRIS spectrum had 
	the slit slightly off-center to avoid a nearby bright star. To correct for the resulting slight flux underestimation after flux calibration the spectrum was scaled to match the reliable SparsePak/WIYN flux level.}
	\label{fig:starspec}
\end{figure}

\section{Observations and Data Reduction}\label{sec:observations}

The stellar remnant 2MASS~J00531123+6730023 has  a {\it Gaia} J2000 position RA: 00$^h$~53$^m$~11.205$^s$, DEC: +67$^o$~30'~02.381'. The {\it Gaia} DR3 parallax is $0.4065\pm0.0259$\,mas and the distance derived from the inverse parallax is $2460\pm 157$ pc. However this transformation is nonlinear and the measured parallax may have large uncertainties for distant and faint objects. In this case, the fractional parallax error is approximately 6.4\%. Therefore in this work, as in \citetalias{paper1}, we adopt the distance measurement of \citet{bailerjones2021} at $2297^{+122}_{-114}$~pc, which was derived from a probabilistic approach on {\it Gaia} data. The Galactic coordinates  ($l$,$b$)\,=\,(123.1, 4.6)  place the stellar remnant at 180\,pc from the plane, so likely part of the old stellar disk.

\subsection{OSIRIS/GTC spectroscopy}

High signal-to-noise spectroscopic data of both star and nebula were obtained with the OSIRIS instrument on the 10-m Gran Telescopio Canarias (GTC) telescope in late 2016 as part of our planetary nebula (PN) follow-up program. The faint nebula spectra from SN~1181 was presented in \citetalias{paper1} with only the [S {\sc ii}] 6716 and 6731\,\AA\ doublet and [Ar {\sc iii}] 7136\,\AA\ emission lines detectable. Here, we focus on the spectra we obtained for the stellar core.

OSIRIS covered the wavelength region from 3700 to 7000\,\AA. The slit width was 0.8\arcsec\ (\citetalias{paper1}). The most prominent stellar feature is the O\,{\sc vi} 3811/34 emission line of remarkable strength below 4000\,\AA\ (Figure~\ref{fig:starspec}), with additional broad spectral features beyond 4200\,\AA. The narrow absorption lines are  known diffuse interstellar bands (DIBs) at 4428, 4502, 4727, 4762, 5418, 5456, 5779, 5796, 6204, 6281, and 6613\,\AA, identified using the NASA catalogue\footnote{\url{https://leonid.arc.nasa.gov/DIBcatalog.html}} \citep{DIBs}. 

\subsection{SparsePak/WIYN fibre spectroscopy}

We analyzed previously un-extracted spectra of the stellar core from the SparsePak fibre-based (fiber diameter 4.7\arcsec), pseudo-integral field unit on the 3.5-m WIYN telescope taken in late 2014. The spectral observations of the surrounding nebula Pa\,30 are presented in \citetalias{paper1}.

\subsection{Kermerrien Observatory spectroscopy}

Low resolution ($R\sim540$) optical spectroscopy of Pa\,30's central star by our amateur collaborators with a 20-cm $F/D\sim5$ telescope at Kermerrien observatory in Porspoder, France, brought its remarkable spectrum and its strong O\,{\sc vi 3811/34 emission line} to our attention in October 2018. The low signal-to-noise spectra were not properly flux-calibrated and were scaled and smoothed to be compared against our OSIRIS/GTC spectrum in Figure~\ref{fig:starspec}.


\subsection{Archival data}

\subsubsection{X-ray observations}\label{sec:xray}

Faint X-ray emission was first recorded by the \emph{ROSAT} All-Sky Survey \citep{rosatfaint} within 12\arcsec\ of the central star at a PSPC count rate in the 0.1--2.4 keV band of $17 \pm 7$ cnts~ks$^{-1}$. Most counts ($>$70\%) have energies greater than 0.5 keV.  
A \emph{Swift} XRT \citep{evans2014} 1.94~ks observation (ID 00358336000, 26$^{\rm th}$ July 2009) serendipitously registered this source.  Analysis of these observations confirms a point source detection at the star's position with a \emph{Swift} XRT count rate in the 0.2--5 keV band of $3.6 \pm 1.4$ cnts~ks$^{-1}$, together with possible diffuse emission. 

Recently \citet{oskinova2020} presented deep, sensitive \emph{XMM-Newton} EPIC X-ray observations that detect a central point source while confirming diffuse nebula emission.  Their spectral analysis of the diffuse emission describes an optically-thin, thermal plasma composed  of carbon, oxygen, neon, and magnesium (see their Table~1). These abundances are attributed to carbon-burning ashes, suggesting Pa\,30 is a supernova remnant. The total mass of the X-ray-emitting nebula, estimated as  $\simeq$0.1 $M_\odot$, strengthens this conclusion.

The point source \emph{XMM-Newton} EPIC spectra are described by a hot plasma at temperatures 1--100 MK. A non-thermal component, that  would reduce the plasma temperature to  $\sim$1--20 MK, cannot be excluded. The EPIC spectra are consistent with the \emph{Swift} and \emph{ROSAT} data, but their much higher quality allow a detailed spectral fit \citep{oskinova2020}. The un-absorbed X-ray luminosity in the 0.2--12 keV band implied by \citet{oskinova2020} is reduced to $6.6\times10^{32}$ erg~s$^{-1}$ at our adopted distance of $2.3\pm0.1$\,kpc.

To contribute SED high energy points presented in Section~\ref{sec:shellphotom} we re-processed the \emph{XMM-Newton} EPIC MOS1, MOS2, and {\it pn} data corresponding to Obs.ID.\, 0841640101 and 0841640201  using \emph{SAS} and relevant calibration files. 
MOS1, MOS2, and {\it pn} background-subtracted spectra of the central star of Pa\,30 were extracted using a circular aperture (radius=20\arcsec) with suitable background regions and calibration matrices computed using the corresponding \emph{SAS} tasks. The 0.2--7.0 keV extended emission image reveals a shell with a diameter of $\approx$4\arcmin\ with a filled morphology. 

\subsubsection{Infrared imaging}\label{sec:obsir}

We mined the IRSA/IPAC archive\footnote{\url{https://irsa.ipac.caltech.edu/}} for available infrared imaging  3\arcmin\ around Pa\,30. Apart from the WISE data presented previously \citep{gvaramadze2019, oskinova2020} and \citetalias{paper1}, usable mid- and far-infrared images exist from AKARI \citep{2015PASJ...67...50D} and  IRAS \citep{1984ApJ...278L...1N}. 
Only the central star is visible in the near-infrared, while the nebula is prominent beyond 10\micron. The far-infrared maps (i.e., AKARI and IRAS) do not have sufficient resolution to resolve the central star from the nebula, but they resolve Pa\,30 which retains its apparent circular shape.  The 12\micron\ WISE image shows a round nebula with a radius of 83\arcsec$\pm$4\arcsec, corresponding to a physical radius of $0.93\pm0.04$ pc, identical to the size of the [O {\sc iii}] shell, \citetalias{paper1}. The extended X-ray shell seen by XMM-Newton is roughly 120\arcsec\ in radius. No nebular emission has been detected in submillimeter {\it Planck} maps.


The infrared photometry from these datasets assume unresolved sources although the nebular component can exceed the size of the apertures. We re-evaluate the photometric measurements taking into account the observed nebular size. In Section~\ref{sec:shellphotom}, we present estimates of the nebula's surface brightness.

\subsubsection{Photometry}\label{phot}

The central star of Pa\,30 is relatively bright, at $G \approx 15.4$ mag. It has been covered in many sky surveys from which we retrieved all available temporal optical and photometric data together, in order to investigate temporal variability (cf. Sect.~\ref{sec:varphot}).

\subsubsection{Spectroscopy}\label{sec:ultra}

In the following sections, we directly compare our OSIRIS/GTC and SparsePak/WIYN stellar spectra with \citet{gvaramadze2019} and \citet{2020RNAAS...4..167G}. These archival long-slit spectra were very kindly provided by these authors.

We also used reduced archival ultraviolet spectra from the Space Telescope Imaging Spectrograph (STIS) on the Hubble Space Telescope (HST) under program GO15864 (P.I. G.~Graefener; epoch: 4 November 2020). 
The spectral coverage was from 1150 to 3150\,\AA, with a slit width of 0.2\arcsec. These spectra provided a comparison against the {\sc cmfgen} models in Sect.~\ref{sec:nlte}.

Finally, we complemented our observations with very recent {\it Gaia} DR3  BP/RP low resolution spectra of $R=30-100 $ \citep{Montegriffo2022} that extends beyond OSIRIS/GTC, i.e., from 3200\AA\ to 
1$\mu$m and covers the \textcolor{blue}{\bf O\,{\sc vi} 3811/34 emission line}. The {\it Gaia} telescope is outside the atmosphere so telluric lines are not present. The spectra were obtained from the UK Gaia Data Mining Platform\footnote{https://www.gaia.ac.uk/data/uk-gaia-data-mining-platform} and are encoded as Hermite functions. The spectra show instrumental distortions around the  O\,{\sc vi} 3811/34 line. Between 4000 and 4500\AA, the feature wavelengths differ by 30\,\AA\ from our ground-based spectra,  possibly due to the response of the Hermite functions to the sharp edge of the O\,{\sc vi} 3811/34 line. The wavelength calibration appears correct for $\lambda>5000$\AA.


\subsection{Extinction}
\label{sect:ext}

\begin{figure}[tbp]
	\centering
		\includegraphics[width=\columnwidth]{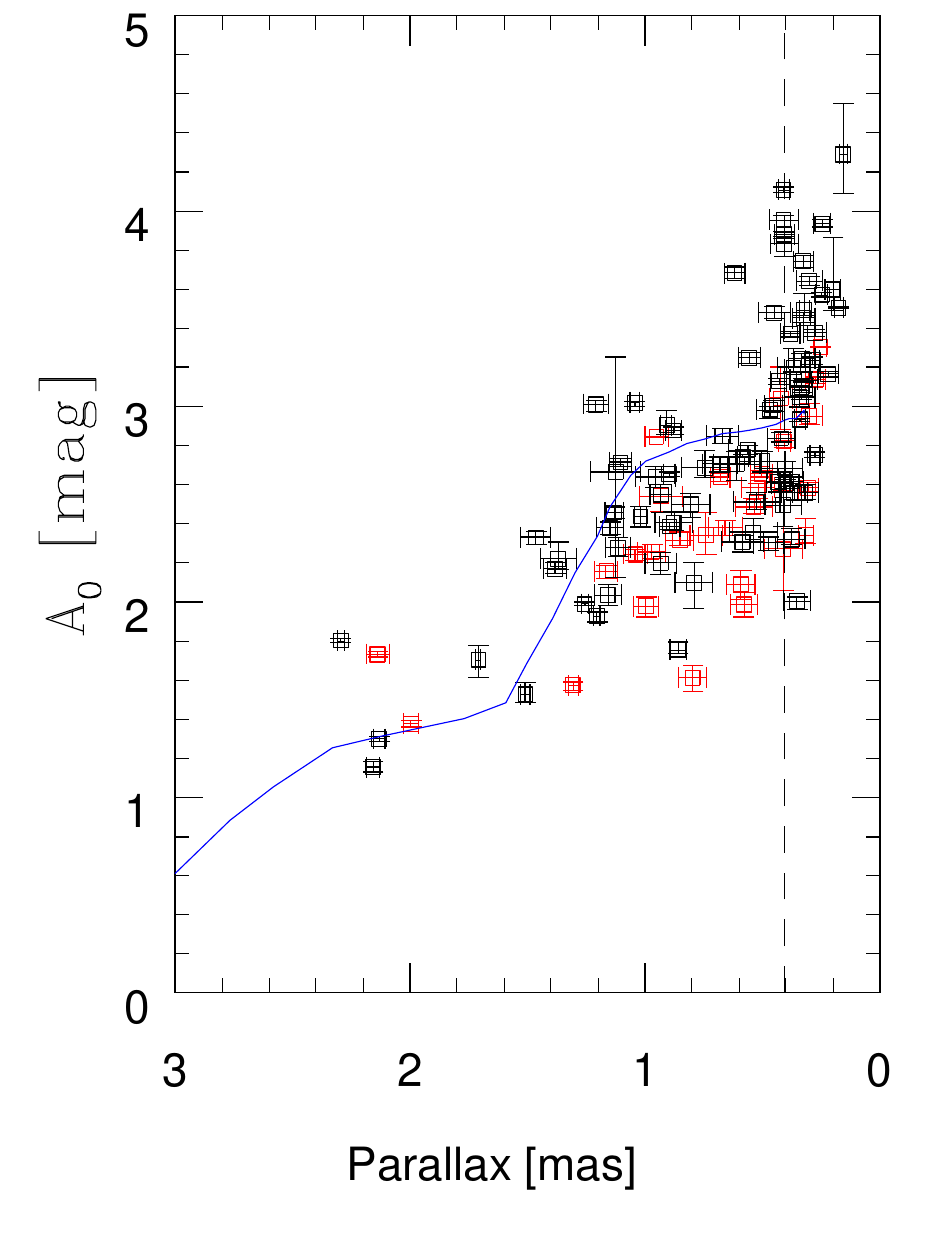}
	\caption{$A_0$ (550\,nm) extinction values for field stars taken from the Gaia DR3 catalogue. Black points are for stars within a radius of 5 arcmin  of Pa\,30, and red points for stars within 2 arcmin. The blue line shows the extinction curve for this direction from \citet{Vergely2022}, and the black dashed line marks the 
	parallax of Pa\,30. }
	\label{fig:extinction}
\end{figure}

\citetalias{paper1} adopted  $A_V=2.4\,\pm\,0.1$~mag but field stars now imply this may be an underestimate. Figure~\ref{fig:extinction} shows parallax and $A_0$ from {\it Gaia} DR3 field stars within a 5~arcmin radius (black points) from the central star of Pa\,30. Red points are stars within 2~arcmin. The curve is the model extinction plot from \citet{Vergely2022} from the {\sc explore} platform\footnote{https://explore-platform.eu/sdas} at a spatial resolution of 25~pc (30~arcmin at the distance of Pa\,30). 
 
At the parallax of the central star (dashed line), the nominal extinction is $A_0=2.9$ mag, albeit with a large scatter between individual stars. The extinction is primarily due to clouds at around 400\,pc and at 800\,pc. The \citet{Vergely2022} curve ends at a distance of 3 kpc. {\it Gaia} DR3 finds some higher extinctions beyond 2.5\,kpc but this is not included in the curve. 
 
The data indicates an extinction $A_0=2.8\pm0.4$ mag. The central value is the same as listed for our star in {\it Gaia} DR3, but that is fortuitous in view of the complexity of the spectrum. In this paper we will argue that the extinction for the central star is at the higher end of the range.
 

\section{The stellar remnant}\label{sec:CS}

\subsection{The stellar spectrum}\label{sec:spec}
\begin{figure*}[htbp]
\gridline{\fig{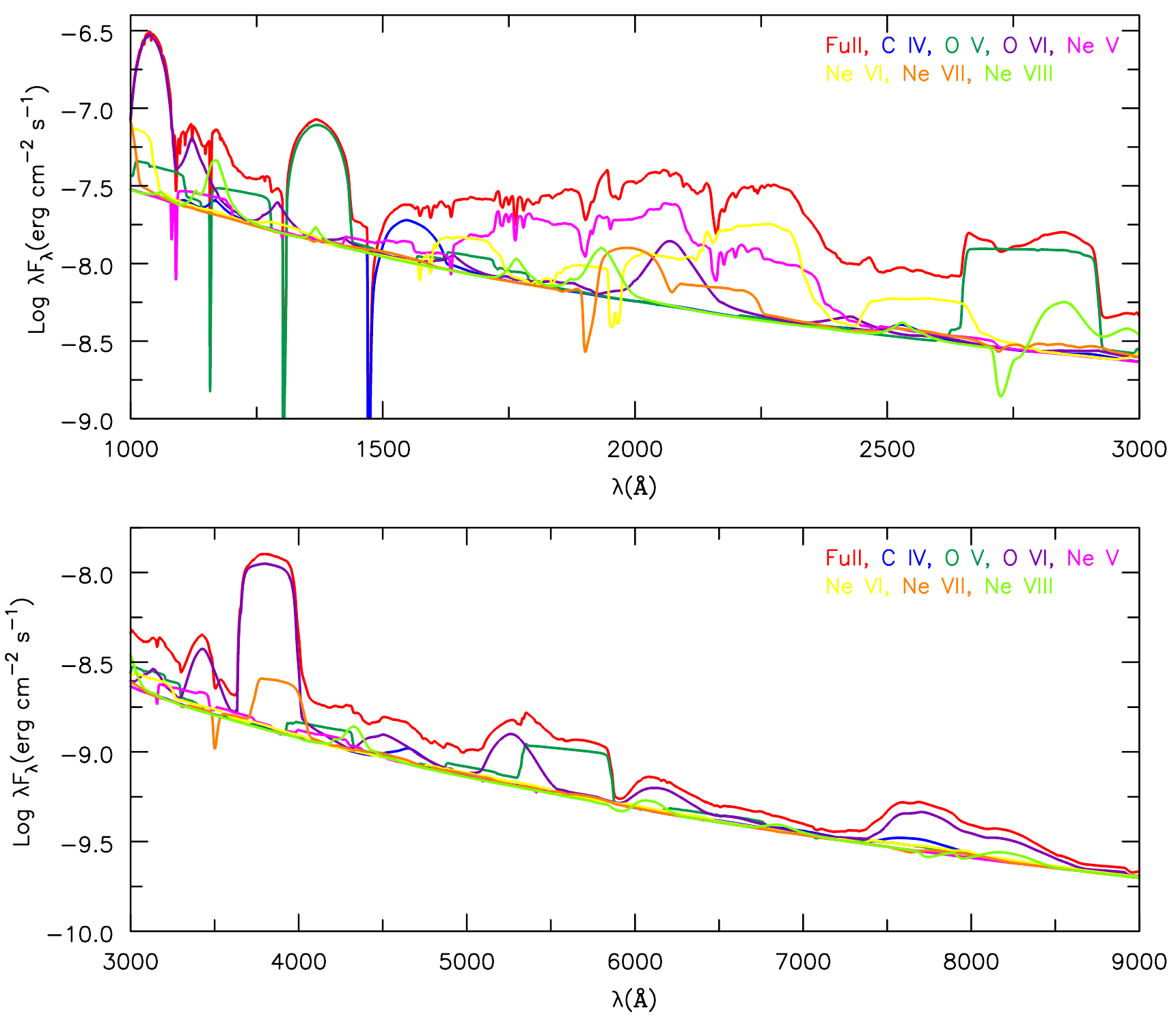}{0.6\textwidth}{(a)}}
\gridline{\fig{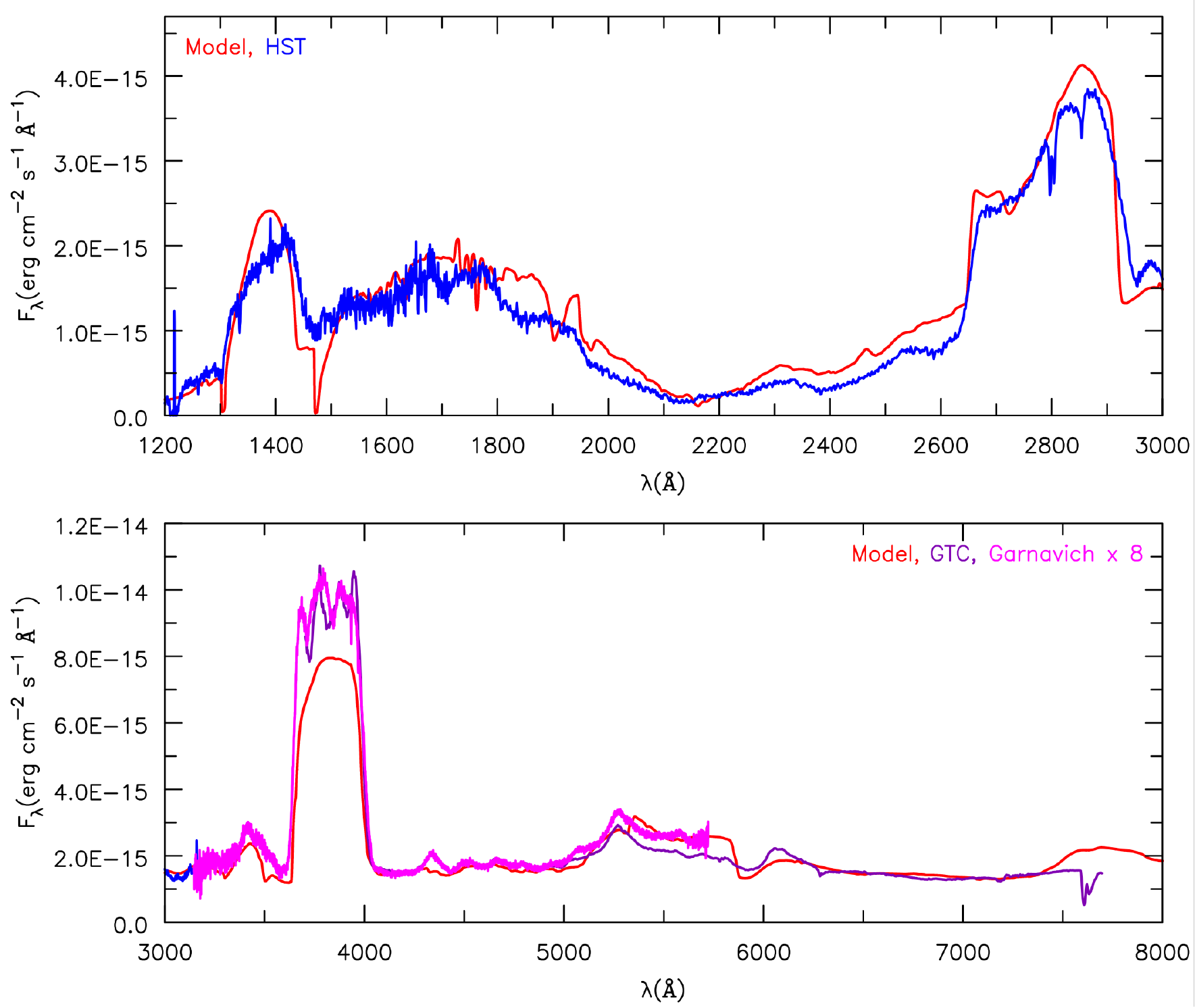}{0.6\textwidth}{(b)}}
	\caption{{\bf (a)} An example of an oxygen- and neon-rich model by {\sc cmfgen}
	(He/C/O/Ne mass fractions :  0.017/0.129/0.414/0.436) where contributions by individual C, O and Ne lines are clearly indicated. The O {\sc v} and O {\sc vi} lines broaden certain spectral features in the model spectrum. Luminosity and temperature was 36,000~L$_{\odot}$ and 237,000 K, respectively. 
 The strongest optical lines are: \ion{O}{6}\ $\lambda\lambda 3811-3833$ (3p\,$^2$P$^{\scriptstyle \rm  o}$ - 3s\,$^2$S), \ion{Ne}{7}\ $\lambda\lambda 3859-3910$ (3s\,3d\,$^3$ D - 2s\,3p\,$^3$P$^{\scriptstyle o})$, \ion{Ne}{8}\ $\sim \lambda 4341$ (9 - 8),
\ion{O}{6}\ $\sim \lambda 5291$ (7 - 6), \ion{O}{5}\ $\lambda\lambda 5572-5604$ (3s\,3d\,$^3$D - 2s\,3p\,$^3$P$^{\scriptstyle o})$, and \ion{O}{6}\ $\sim\lambda 7737$ (8 - 7). {\bf (b)} The {\sc cmfgen} model (red) extended to the ultraviolet and compared to archival STIS/HST spectra (blue), the OSIRIS/GTC spectrum (purple), and the 
\cite{2020RNAAS...4..167G} spectrum (scaled, pink).
	}
	\label{fig:model_n38}

\end{figure*}

The central star of Pa\,30 has a unique spectrum (Figure~\ref{fig:starspec}) with  very broad, very high excitation emission lines and a striking discontinuity bluewards of 4000\,\AA. There is a complete absence of hydrogen, helium and  nitrogen lines. The broad lines are mostly identified with very high ionization stages of oxygen (Figure~\ref{fig:model_n38}). The sharp and significant emission line that cuts-on around 4000\,\AA\ is identified as O {\sc vi} 3811/34\,\AA. This feature turns over near the atmospheric cut-off as confirmed in the recent spectrum of \cite{2020RNAAS...4..167G}. This doublet is the defining criteria for the WO spectral type in both Pop~I stars \citep[e.g.,][]{crowther1998, gvaramadze2019}, as well as old low-mass central stars of planetary nebulae (designated as [WO]) before they enter the white dwarf phase \citep[e.g., ][]{smith1969, sanduleak1971}. The broadened emission at 5270/91\,\AA\ is identified as O {\sc vi}. 
The central star of Pa\,30 seems unique in arising from neither pathway if the double degenerate merger and accretion fed Type~Iax SN are correct \citep{gvaramadze2019,oskinova2020}.

In addition to these strong, broad features, a number of weaker, individual lines are seen. Their profiles can be accurately fitted with Gaussians with FWHM of $0.015c$. The wings extend to velocities as high as 15,000\,km\,s$^{-1}$ ($\approx0.05c$). Several lines coincide with  O~{\sc viii} and are tentatively identified as such.  They are well centered on the systemic velocity. 

\subsection{The stellar wind {\sc cmfgen} models}\label{sec:nlte}

The broad, prominent spectral features are indicative of a strong stellar wind. We modeled it using the non-LTE line-blanketed radiative transfer code {\sc cmfgen}\footnote{\url{http://kookaburra.phyast.pitt.edu/hillier/web/CMFGEN.htm}} adapted to WC and WO stars \citep{hillier1998,hillier2012a,hillier2012b}. Input parameters were our OSIRIS/GTC spectrum that includes the O {\sc vi} doublet at 3811/44\,\AA\ with the interstellar extinction derived in \citetalias{paper1}, and a wind velocity of 15,000\kms. A classic velocity law is used,

\begin{equation}
v(r)=v_{\infty}\left( 1-{r\over R_*}\right),
\end{equation}

\noindent where $v_{\infty}$ is the wind terminal velocity measured from the spectra and $R_*$ is the stellar radius. There is a consensus that stellar winds are clumped. This is taken into account in {\sc cmfgen} modeling via a volume filling factor, adopted here as $f=0.1$. For the spectral modeling the ratio $\dot{M}/\sqrt{f}$, where $\dot{M}$ is the mass-loss rate, can be considered an invariant. 

We explored a range of stellar luminosity and temperature, mass-loss rate, stellar radius, velocity law and chemical abundances, summarized in Table~\ref{tab:models}. Given the limited spectral coverage of our OSIRIS/GTC spectrum (and indeed of all available optical spectra- except that from Gaia, see above), we could not find a unique model that completely reproduces the observations \citep[unlike ][although they also stated large uncertainties, e.g., Table~\ref{tab:models}]{gvaramadze2019}. Nevertheless, many models could reproduce most spectral features. The results shown are best approximations and/or upper limits to the individual parameters. Most models tested could reproduce the visual and near-infrared photometry. Determining errors on the fits is tricky and we refrain from giving error bars as the errors are not easily transformed into a standard deviation and strong coupling exists between some parameters. Potentially four species can have mass fractions $>$0.1, so changing the mass fraction of one species necessitates a corresponding change of at least one other species. Therefore, below, we provide viable ranges for the stellar 
parameters and indicate the properties used to constrain them.

The three primary modeling constraints come from matching the mean flux level, the strength of the O\,{\sc vi} 3811/34\,\AA\  doublet, and the O\,{\sc v} 5572--5604\,\AA\ line strength.  Only models with luminosities between 25,000 and 60,000 L$_\odot$ provide reasonable fits. Below the low luminosity limit either the O\,{\sc vi} doublet is too weak and/or O {\sc v} too strong, while at the upper limit either O\,{\sc vi} is too strong and/or O\,{\sc v} is too weak. At very high luminosities the  O\,{\sc vi} doublet strength is suppressed (but other O\,{\sc vi} lines remain strong) since the upper levels preferentially decay directly to the ground state. We hence set an upper limit at $\leq$260,000~K\footnote{This is the classical effective temperature defined at a Rosseland optical depth of 2/3. Due to the dense wind this is somewhat lower (around 10\%) than would be defined using the radius of the star at the sonic point.}.

The peak near 5300\,\AA\ is due to O\,{\sc vi} offset to the red from observations, perhaps due to the overlapping O\,{\sc v} line being too strong in the model (Figure~\ref{fig:model_n38}). 
The model matches bumps seen at $\sim$4500\,\AA\ and $\sim$4659\,\AA, but fails to predict the feature near 4250\,\AA\ which is assigned to O\,{\sc viii}. An alternative suggestion for this emission is Ne\,{\sc viii}, which has approximately the same wavelength as the O\,{\sc viii} 
line. In a model with a slightly lower mass loss, the Ne\,{\sc viii} is clearly visible in predicted spectra though the overall fit to the spectrum is worse. As apparent from Figure~\ref{fig:model_n38}, the broad emission lines blend together making it impossible to see the true continuum.

We provide some upper limits for mass fractions of key elements (He, C, O and Ne) assuming Solar abundance ratios for other elements, that is Na, Mg, Al, Si, S, Ar, Cl, Ca, Fe, and Ni\footnote{f For the following transitions: Na\,{\sc iv} - Na\,{\sc x}, Mg\,{\sc iv} - Mg\,{\sc xi}, Al\,{\sc iv} - Al\,{\sc xii}, Si\,{\sc iv} - Si\,{\sc xi}, S\,{\sc v} - S\,{\sc xi},  Ar\,{\sc vi} - Ar\,{\sc x}, Cl\,{\sc vi} - Cl\,{\sc viii}, Ca\,{\sc vi} - Ca\,{\sc xii}, Fe\,{\sc vi} - Fe\,{\sc xvii}, and Ni\,{\sc vi} - Ni\,{\sc xvii}}. In these hot models helium has little influence on the spectra. From the apparent absence of the He\,{\sc ii} 4686\,\AA\ line we place an upper limit of 0.4 on $X_{\rm He}$. The O\,{\sc vi} features at 3811/34\,\AA\ and 5270/91\,\AA\ are clear indicators that the stellar atmosphere is oxygen-rich. Therefore we used an oxygen-rich model atmosphere with an oxygen mass fraction $<0.647$. This reproduces the O\,{\sc vi} doublets, as well as O\,{\sc v} lines. All the strongest features can be reproduced but not some weaker individual ones. The lines at 4500 and 4658\,\AA\ are sitting on a broader O\,{\sc vi} feature whose strength depends on continuum choice, uncertain due to the density.

Carbon-free and carbon-enhanced models were tested.  The C\,{\sc iv} line can contribute to the 4568\AA\ line but otherwise there is no easily distinguishable carbon line contribution.  We set an upper limit for $X_{\rm C}\leq0.261$.  We explored a range of neon abundances 
(Table~\ref{tab:models}) and find an enhanced neon abundance could contribute to the spectral features at the 3600--4000\,\AA\ plateau (Ne\,{\sc vii}), at 4341\AA\ and at 6070\AA\ (Ne {\sc viii}). Inclusion of neon does not affect our ability to get a good fit to the oxygen lines. On the other hand, the 4341\,\AA\ line can also be  identified with O {\sc viii} (see below).

The fit shown in Figure~\ref{fig:model_n38} provides a `maximum' model where as many features as possible are reproduced. Oxygen is well fitted, but carbon may be less certain and neon is likely significantly over predicted. The plotted model has $X_{\rm Ne} =0.436$ and may be considered as the upper limit. A more carbon-rich composition cannot reproduce the spectrum. The ranges of chemical abundance of carbon, oxygen and neon in Table~\ref{tab:models} agree with those derived by \citet{oskinova2020} from the central star's X-ray spectrum.

\begin{table*}[htbp]
\centering
	\caption{Range of stellar parameters for {\sc cmfgen} models. The last two columns show the values from \citet{gvaramadze2019} and from the hot gas model in the Sect.~\ref{sec:cloudy}. 
	}
 \label{tab:models}
	\begin{tabular}{lccccc}
	\hline
	Parameter & Explored range & Best fit range & Illustrated model & \citet{gvaramadze2019} & 
	Hot gas model \\
	\hline
$L_*$ ($\rm L_{\odot}$) & 10,000 -- 200,000  & 30,000 -- 60,000 & 36,000 & 
$39,810^{+20,144}_{-10,970}$ & \\
$T_*$ (K) & 145,000 -- 580,000 & 200,000 -- 280,000   &  237,000 & 211,000$^{+40,000}_{-23,000}$ & 
\\
$\dot{M}$ ($\rm M_{\odot}$/yr) & 7.5$\times$ 10$^{-7}$ -- 2.5$\times$ 10$^{-6}$ & $\leq$ 4$\times$ 
10$^{-6}$ & 2.6 $\times$ 10$^{-6}$ & $3.5(\pm0.6)\times 10^{-6}$ & \\
$R_*$ ($\rm R_{\odot}$) & 0.04 -- 0.22 & $\leq 0.2$ & 0.155  & 0.15$\pm$0.04 & \\
v$_{\infty}$ (kms) & -- & $\sim$15,000 & 15,000 & 16,000$\pm$1,000 & \\
\hline
	Mass fractions & & & \\
	\hline
$X_{\rm H}$ &  --  & -- & -- & -- & -- \\
$X_{\rm He}$ &  0.017 -- 0.135 & $<0.4$ & 0.017 & $<0.1$ & $\leq 0.44 $\\
$X_{\rm C}$ &  0.0 -- 0.261 & $\leq 0.26$ & 0.13 & $0.2\pm0.1$ & 0.13 \\
$X_{\rm O}$ &  0.414 -- 0.697 & $\leq 0.7 $ & 0.41 &  $0.8\pm0.1$ & 0.39  \\
$X_{\rm Ne}$ & 0.011 -- 0.501 & $\leq 0.5 $ & 0.44 & 0.01 & 0.04 \\
\hline
	\end{tabular}
\end{table*}

The broad wind profile ($\approx$15,000\kms) provides a slightly better fit to the broad spectral
profiles for both the O\,{\sc vi} doublets and O\,{\sc v} feature at 5900\,\AA\ 
(Figure~\ref{fig:model_n38}). This value is close to the measured value of \citet[][16,000\kms]{gvaramadze2019}. Since O\,{\sc v} is present, this emerges from the outer regions, and so a higher terminal velocity is needed to fit both oxygen features simultaneously. As found by \cite{gvaramadze2019} radiation pressure appears insufficient to drive the wind. Though the OSIRIS/GTC spectra do not extend below 3600\,\AA, the {\sc cmfgen} models predict an O\,{\sc vi} line at 3435\,\AA\ similar to \citet{gvaramadze2019} and has been observed in the spectra of \cite{2020RNAAS...4..167G}.

After finalizing the model, we compared it to HST ultraviolet (UV) spectra of the stellar remnant (Sect.~\ref{sec:ultra}). To fit the UV data we needed to adjust the reddening to E(B-V)$=1.05$. Although we eventually updated our reddening, our initial models started off assuming the reddening from \citetalias{paper1}. With the new reddening the fluxes no longer matched, so we recomputed the model with scaled parameters that would yield the same spectral shape, and this model is shown in Figure~\ref{fig:model_n38}.\footnote{
To a good approximation an accurate distance is not needed for the spectral fitting. For stars with dense stellar winds, models with the same abundances, effective temperature, and value of $\dot{M}/R_*^{3/2}$ produce similar spectra --  only differing in flux \citep{1989A&A...210..236S}. The parameters for one distance can be scaled to a new distance using the following scaling relations: $L \propto d^2$, $R_* \propto d$ and $\dot{M} \propto d^{3/2}$.}  Given the O\,{\sc vi} 3811/34 emission line's strength, the estimated reddening should be more accurate than that obtained using optical fluxes over a limited wavelength range.

No further model modifications were made and there is excellent agreement between the STIS/HST fluxes and the OSIRIS/GTC spectrum. Figure~\ref{fig:model_n38} shows the {\sc cmfgen} model correctly predicts the O\,{\sc v} and Ne\,{\sc viii} blend at $\sim$2800\,\AA\ and the strong O\,{\sc v}  emission at $\sim$1370\,\AA. The model also predicts strong O\,{\sc vi} resonance emission (1032, 1038\,\AA). Absorption shortward of 1500\,\AA\ is from the C\,{\sc vi} resonance transition (1548, 1551\,\AA) being too strong, possibly indicating a lower carbon abundance than in the model. The strong 2200\,\AA\ band and its Galactic variation make it difficult to judge agreement between model and observation for features in the band. 


We computed additional models to better constrain the parameters and potentially improve the fit, but this did not lead to a change in the adopted parameters. One possible exception is the neon abundance -- a model  with the abundance reduced by a factor of 2  might  give a better fit  to the UV spectrum. However, a better understanding of the interstellar 2200\,\AA\ band is required to draw firm conclusions.

In a recent study of two WO stars in the Large Magellanic Cloud by \cite{Aadland2022} it was found that X(C)/X(O) was greater than 5  despite the presence of strong  O\,{\sc vi} features at 3811/34\AA. However this star is descended from a massive O star and has a very different evolutionary history than the central star of Pa30. Further on, the
spectra are quite distinct - in the WO stars, for example, the C\,{\sc iv} spectrum is still very prevalent and neon lines are relatively weak and can be matched by a neon abundance similar to that expected from the processing of CNO into $^{14}$N in the CNO cycle, and the subsequent processing into $^{22}$Ne. The high temperatures used in our models will weaken the C\,{\sc iv} lines, but this alone cannot explain the weakness of the carbon spectrum and the strength of the oxygen spectrum.

\subsection{Shocked gas: {\sc cloudy} models  }\label{sec:cloudy}


The X-ray spectrum indicates an additional high temperature gas component \citep{oskinova2020} that is strong compared to what is typical for OB stars.  The  unabsorbed X-ray to bolometric luminosity ratio, $\log(L_{\rm X}/L_{\rm bol})$, is in the range of $-4.8$ to $-6.1$, one to two orders of magnitude larger than OB stars, i.e., $\log(L_{\rm X}/L_{\rm bol}) = -6.912\,\pm\,0.153$ \citep{Sana2006}. \citet{oskinova2020} attribute the bulk of the remnant's thermal X-ray emission to the wind's outer regions. 

The wind model describes well the shape and many features of the optical spectrum. However, some  weaker lines are not fitted and the 4340\,\AA\ feature requires Ne\,{\sc viii} at a high neon abundance.  This line coincides with a high-excitation O\,{\sc viii} line. We therefore explore a spectral contribution from the high-temperature gas component which is not included in the wind model.

As a hydrogenic ion, O\,{\sc viii} has a restricted number of  lines. The 4340\,\AA\ line coincides with the O\,{\sc viii} 8--9 transition and the 6069\,\AA\ line with the O\,{\sc viii}  9--10 transition. These transitions have a wavelength ratio of 5/7 and the two line profiles overplot accurately when wavelengths are scaled by this ratio. The observed intensity ratio of the two lines (after subtraction of an estimated continuum) is reproduced by these O\,{\sc viii} lines for an extinction in agreement with the foreground extinction. This strengthens identification as O\,{\sc viii}, although other lines may contribute to these features. The temperature required for O\,{\sc viii} is $>$1\,MK, far above the photospheric or wind temperature.

We ran exploratory photo-ionisation models using the {\sc cloudy} code \citep[ver. 17.03][]{cloudy}.  In the `coronal mode' gas temperature is pre-defined and all lines are assumed collisionally excited. There is no input radiation field. A single shell model was used, with a constant density and temperature and without considering optical depth effects. Models were run  for a range of temperatures and abundances. The hydrogen density was set to $\log n = 7.5$.  All lines were assumed to have a Gaussian profile with sigma of $v/c=\Delta \lambda/\lambda = 0.0092$ as found from line fitting.  The {\sc cloudy} line output list was convolved with this Gaussian profile to obtain model spectra.  The model spectra were multiplied by the extinction curve of \citet{ccm2} and scaled to the peak intensity of the 4340\,\AA\ line. The ratio of the integrated line strength of the 4337\,\AA\ and 6064\,\AA\ lines was used to derive the extinction $A_V$, where all contributions to these features were taken into account. 

The broad wind features at 3800\,\AA\ and 5200\,\AA\ are not reproduced in this model: the temperatures are too high for significant O\,{\sc vi} emission.  However, the weaker lines absent from the wind model could be reproduced with this hot gas.

\begin{figure}[tbp]
	\centering
		\includegraphics[width=\columnwidth]{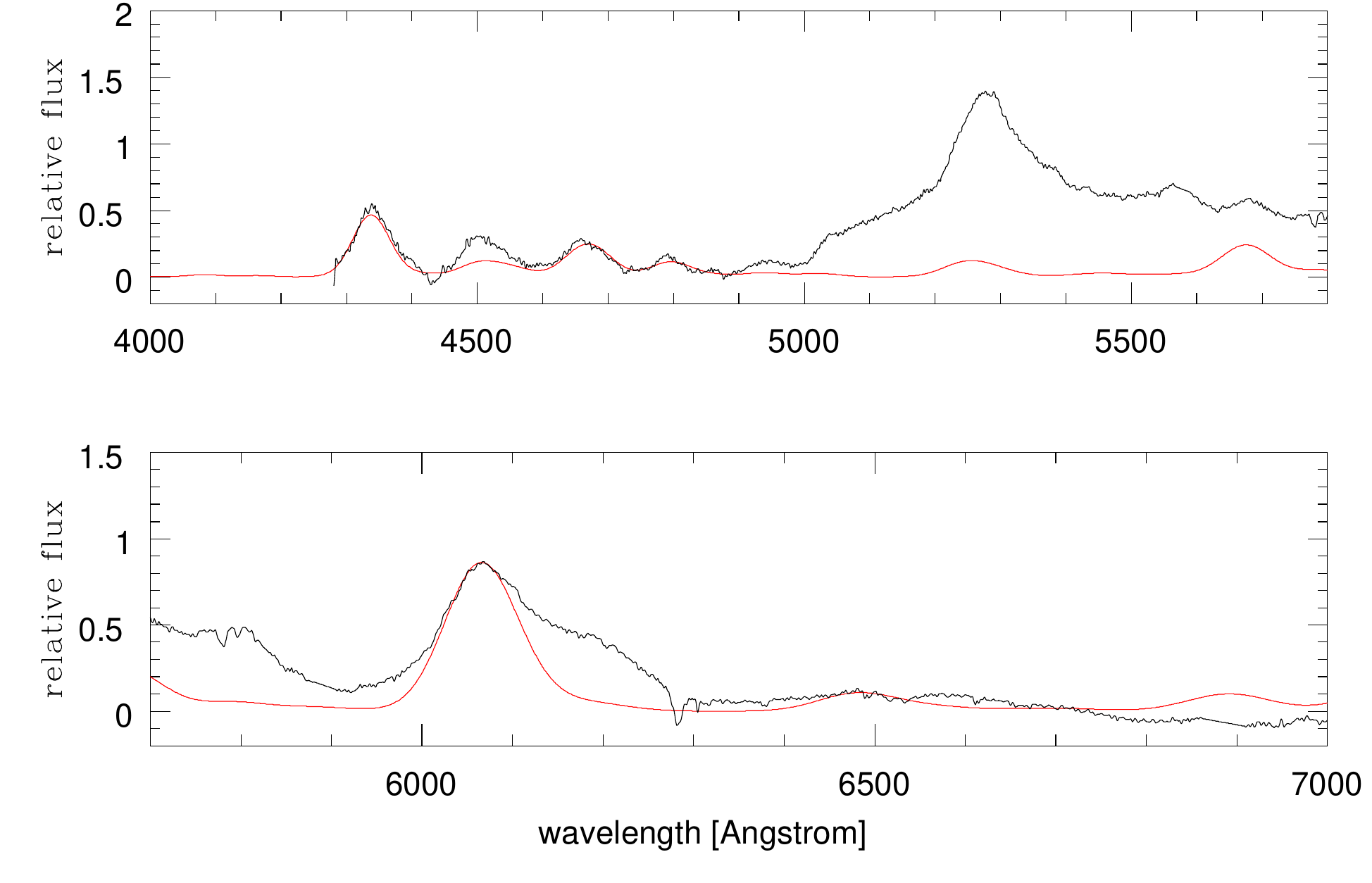}
	\caption{A {\sc cloudy} model for the hot gas emission at a gas temperature of 4~MK (red line) and $A_V=3.0$ mag compared to the central star's SparsePak/WIYN spectrum (black) . The spectrum has  an estimated constant continuum level subtracted of 1.45 on the scale of the plot. For further description see Sect.~\ref{sec:cloudy}. }
	\label{fig:cloudycorona}
\end{figure}

The model fit is shown in Figure~\ref{fig:cloudycorona} along with the observed spectra for a gas temperature of $T=4$\,MK. The best fit gave abundances  (number ratios) of He/C/N/O/Ne = 0.73/0.07/0.02/0.16/0.01, where helium must be considered as an upper limit. These abundances are within the wind model range. The brightest predicted lines between 3000\,\AA\ and 7000\,\AA\ are listed in Table \ref{tab:lines}. 

An estimated constant-value continuum was subtracted from the observed spectrum across the full wavelength range with no attempt to correct for any continuum slope. The line strengths depend on the continuum choice. This introduces an uncertainty, as the dominant broad wind features obliterate most of the continuum. The subtraction of a constant level from the WIYN spectrum brought both the blue and red ends to zero. For the GTC spectrum, a constant subtraction left an offset between the ends. We used the WIYN spectrum for the fit. The procedure indicated an extinction of $A_V=2.9\pm0.2$ mag from the WIYN data, in excellent agreement with that found in Sect.~\ref{sect:ext} and with the UV extinction indicated by the wind model. We adopted $A_V=3.0$ mag for the final model. The model is shown with the continuum-subtracted WIYN spectrum in Figure \ref{fig:cloudycorona}.

\begin{table}[hbtp]
	\centering
	\caption{Predicted, brightest ultra-high excitation (UHE) lines from the {\sc cloudy}  models. The line intensity, $I_0$, is relative to the 4337\,\AA\ O {\sc viii} 8-9 transition, and are not corrected for extinction. }\label{tab:lines}
	\begin{tabular}{clc}
	\hline
	$\lambda$ (\AA) &	 Ion &	 $I_0$ \\
	\hline
4337 &	 O\,{\sc viii} 8-9   & 1.00	 \\ 
4498 &	 O\,{\sc viii} 11-14 & 0.12	 \\ 
4655 &	 O\,{\sc viii} 10-12 & 0.24  \\ 
4681 &	 O\,{\sc viii} 12-16 & 0.06	 \\ 
4686 &   He\,{\sc ii} 3-4    & 0.07   \\
4795 &	 Ne\,{\sc ix}        & 0.29	  \\ 
4940 &	 Ne\,{\sc x} 12-14   & 0.05    \\ 
5243 &	 Ne\,{\sc x} 10-11   & 0.16    \\ 
5288 &	 C\,{\sc vi} 7-8     & 0.05    \\ 
5670 &	 O\,{\sc vii}        & 0.16	  \\ 
5690 &	 O\,{\sc viii} 12-15 & 0.08	  \\ 
6060 &	 O\,{\sc viii} 11-13 & 0.15	  \\ 
6064 &	 O\,{\sc viii} 9-10  & 0.54	  \\ 
6481 &	 Ne\,{\sc ix}        & 0.13	  \\ 
6894 &	 Ne\,{\sc x} 10-11   & 0.09    \\
7080 &  O\,{\sc viii} 13-16  & 0.05 \\
7728 &  O\,{\sc viii} 12-18  & 0.09 \\
8203 &  O\,{\sc viii} 10-11  & 0.30 \\
8521 & Ne\,{\sc ix}          & 0.04 \\
8672 &  O\,{\sc viii} 14-17  & 0.03 \\
8857 & Ne\,{\sc x}           & 0.04 \\
9668 &  O\,{\sc viii} 13-15  & 0.06 \\
	\hline	
	\end{tabular}
\end{table}

The {\sc cloudy} model fits the 4337, 4655, and 6064\,\AA\  with the O\,{\sc viii} 8-9, 10-12, and 9-10 transitions, respectively, though the observed 4655\,\AA\ line is stronger than predicted.  Another O\,{\sc viii} line at 5790\,\AA\ coincides with a peak inside the broad wind band. This identification is not as secure because of the unknown optical depth in this wind band.
The 4800\,\AA\ feature is reproduced with Ne\,{\sc ix}. 
The He\,{\sc ii} line at 4686\,\AA\ contributes to the observed line at this wavelength together with the O\,{\sc viii} line. A good fit requires a helium abundance ratio of He/O\,=\,4 (by number). However, there is also a C\,{\sc vi} and an Ne\,{\sc x} line at this wavelength which could contribute. Hence, this helium abundance is considered as an upper limit. The 4940\,\AA\ feature is not reproduced. A weaker model line predicted at 5290\,\AA\ is due to C\,{\sc vi} but it is likely obscured by the broad wind feature.  

Our {\sc cloudy} hot-gas model is able to reproduce most lines not present in the wind model, using abundances equivalent to that of the wind model, for a gas temperature of $T=3.5$--$4$\,MK. Importantly, all lines which are predicted to be detectable are seen in the spectrum, with a sole exception the Ne\,{\sc x} line at 6900\,\AA, which is affected by a telluric feature.

\begin{figure}[tbp]
	\centering
		\includegraphics[width=\columnwidth]{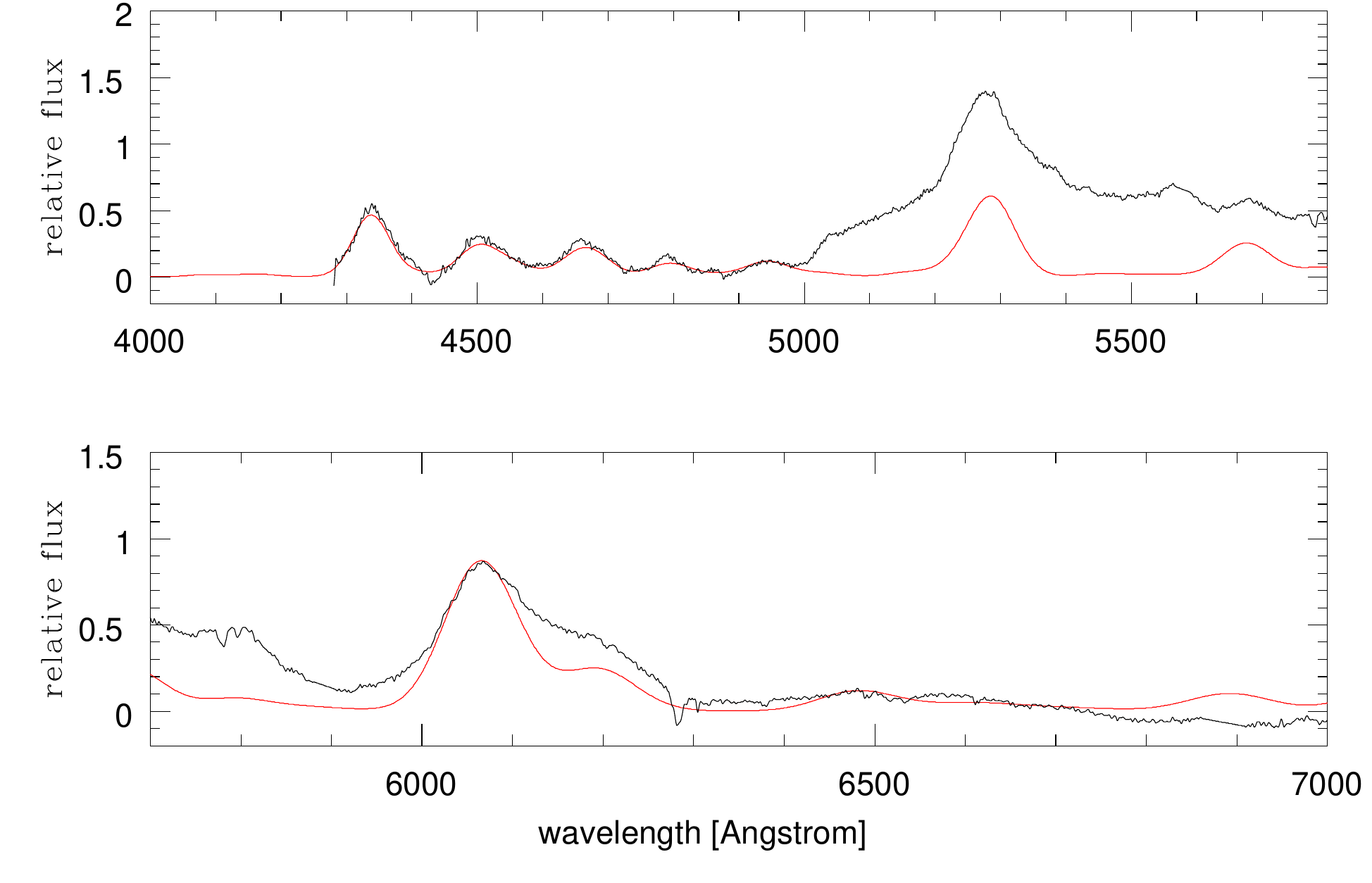}
	\caption{As in Figure \ref{fig:cloudycorona}, but with a much higher carbon abundance and a lower helium abundance. }
	\label{fig:cloudycorona2}
\end{figure}

The carbon abundance is poorly constrained because of a lack of bright lines though several fainter carbon lines do coincide with features in the spectra. Figure~\ref{fig:cloudycorona2} shows a model with the carbon abundance (by number) increased to C/O\,=\,4.5, and with a lower helium abundance. This produces a good fit to many of the lines but adds a strong component to the 5290\,\AA\ line which is already well fitted in the wind model without this component. There is no other evidence for such a high carbon abundance and the wind model needs oxygen-rich gas.  We therefore do not adopt these abundances.

\begin{figure}
    \centering
    \includegraphics[width=\columnwidth]{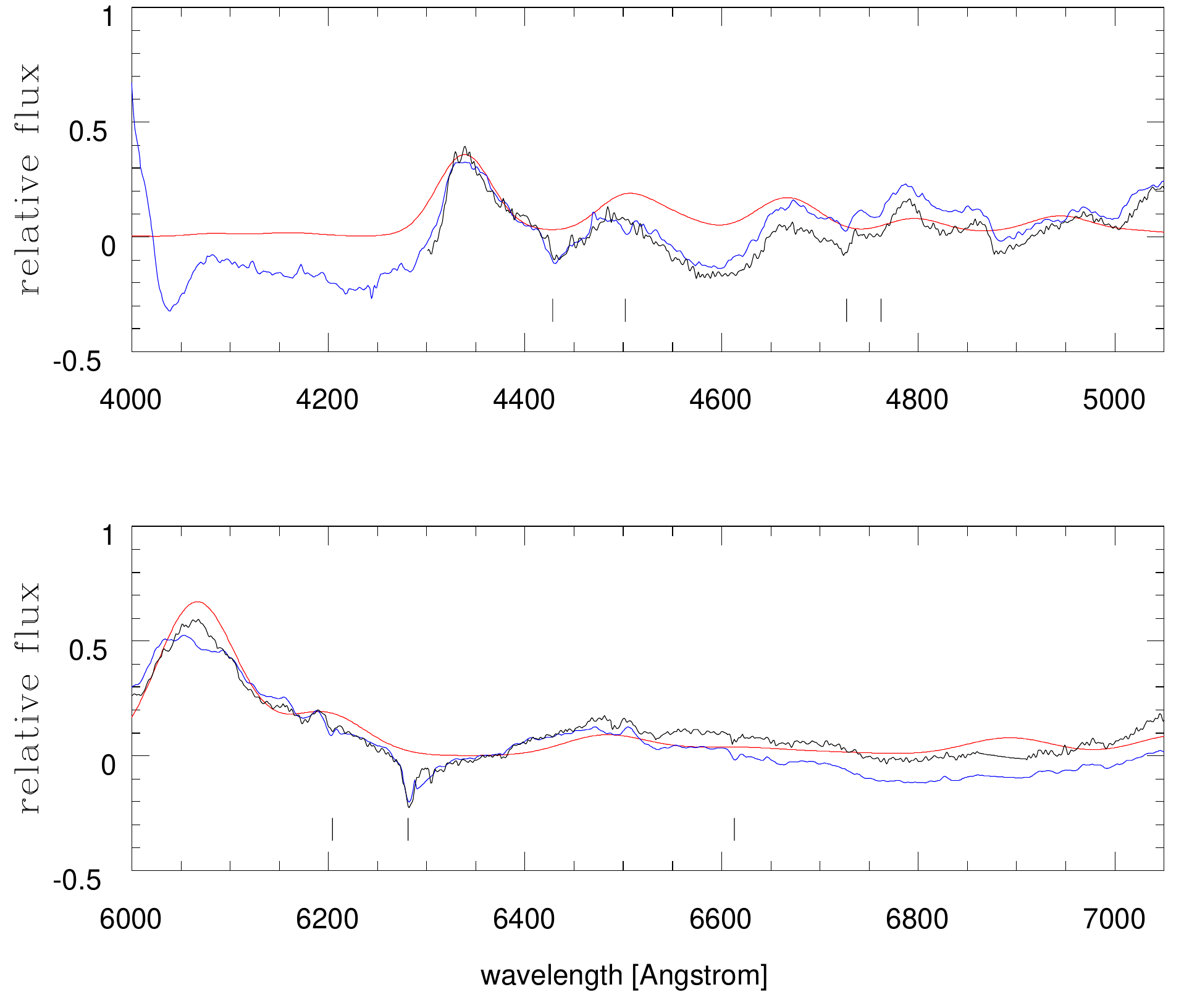}
    \caption{Hot-gas {\sc cloudy} model (red) against the residual SparsePak/WIYN (black) and OSIRIS/GTC (blue) spectra after subtraction of the wind model. A residual continuum of 0.1 for the WIYN spectrum and 0.2 for the GTC spectrum was subtracted shortward of 5000\,\AA. Vertical bars indicate wavelengths of diffuse interstellar bands which are visible in the spectra. The {\sc cloudy} model was scaled to the 4363\,\AA\ line with an extinction $A_V=3.0$.  }
    \label{fig:cloudy_wind}
\end{figure}

Figure~\ref{fig:cloudy_wind} shows the same fit, but now the observed spectra are shown with the wind model subtracted. Only wavelength regions free of the broad wind features are shown. The wind model accounts for almost all of the continuum present at these wavelengths. A small excess was subtracted shortward of 5000\,\AA. This shows that the combination of the wind and hot gas models can explain much of the observed spectra.   

The abundances in the {\sc cloudy} model were adopted from the wind model. The nitrogen abundance is not constrained in this model because of a lack of bright lines. There are potential C\,{\sc vi} lines at 6195\,\AA\ and at 4683\,\AA, which can fit lines at these locations, however these lines need a C/O ratio well above unity to fit the corresponding features, while the wind model does not support C/O\,$>1$. In addition, the 5290\,\AA\ C\,{\sc vi} line  becomes too strong at higher carbon abundance. The temperature is mainly constrained by O\,{\sc viii} lines which affect the 6064\,\AA\ line at temperatures below 3\,MK, and the Ne {\sc x} line at 6894\,\AA\ which becomes too strong for $T>4$\,MK. Because of the model's simplicity these constraints should be viewed with caution, along with the assumption of uniform temperature and density, and the fact that other wind lines may be present.

The line widths and temperature indicate that the hot gas is closely related to the stellar wind. The line widths are similar or slightly smaller to that of the wind. The assumption of a gaussian profile is not identical to the wind profile of Eq (1).  The symmetric line profiles  favor a physical location embedded in the outer wind. The gas can be heated by shocks as regions with different speeds collide. This is a common situation in OB stellar winds.  A smaller line width for the hot gas could argue for a location further inside the wind but this remains to be investigated. 

For Wolf-Rayet winds (hydrogen poor), the shocks can be considered approximately isothermal \citep{Hillier1993}. We use the relation \citep{Hillier1993}:
\begin{equation}
    T_x \approx 5.8 \times 10^6\, {\mu \over 1+\gamma} \left( {v_{\rm s} \over 500\,{\rm km\,s^{-1}}} \right)^2 ,
\end{equation}
\noindent where $v_{\rm s}$ is the differential velocity between wind and shock front, $\mu$ is the mean ionic weight, and $\gamma$ the ratio of electrons to ions. Putting in O\,{\sc viii} as the dominant ion, the second term is approximately unity. For $T_x \sim 5\,$MK,  $v_{\rm s} \sim 500\,\rm km\,s^{-1}$. This is much less than the wind velocity but plausible as internal differential velocities. 

\begin{figure}[tbp]
	\centering
		\includegraphics[width=\columnwidth]{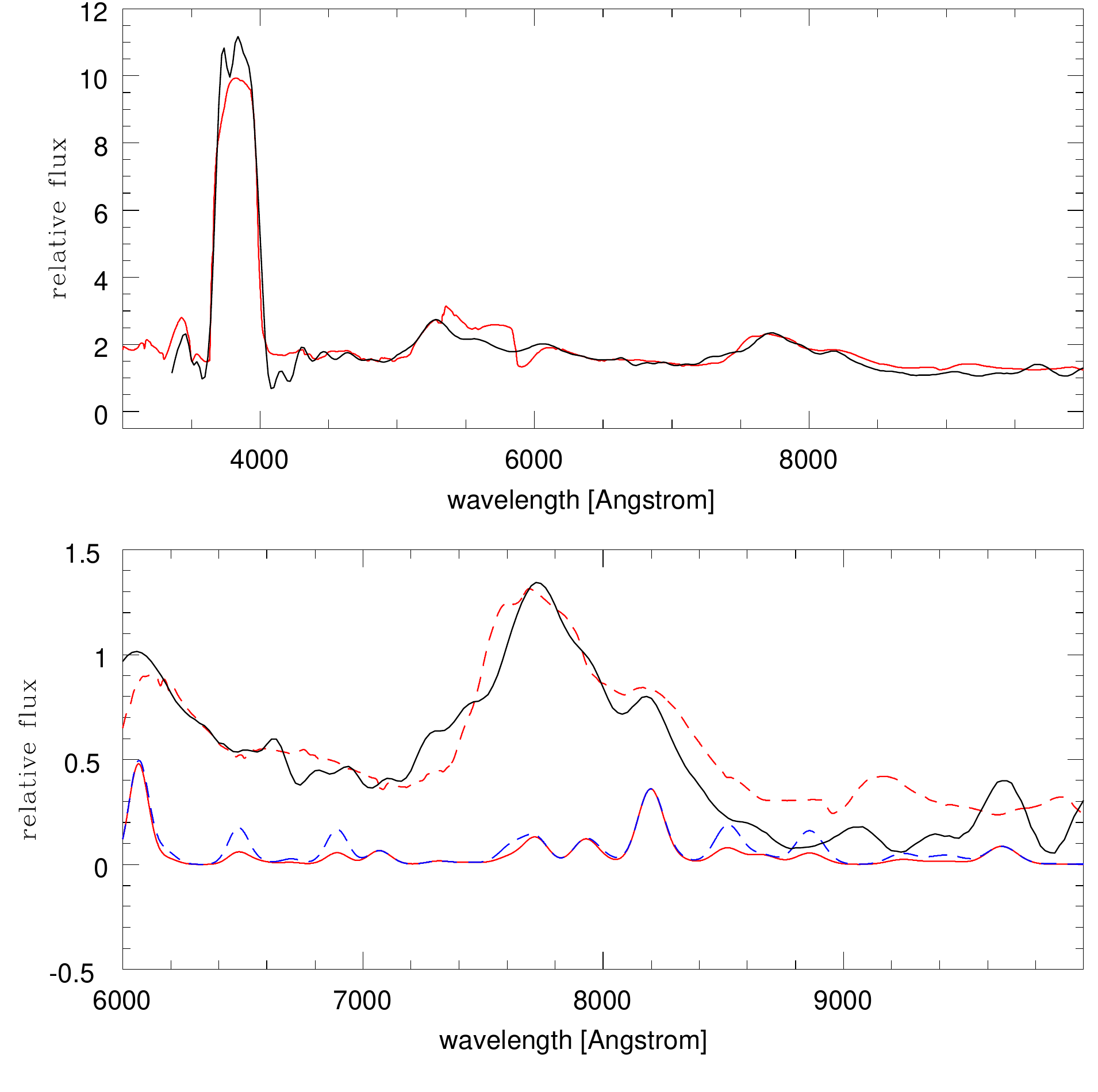}
	\caption{The {\it Gaia} BP/RP spectrum (black). Upper panel: the wind model (red). Lower panel: wind model (dashed, red), the hot-gas model (red), hot gas model with three times higher neon abundance (dashed, blue). In the lower panel, for clarity a level of 1.0 has been subtracted from the {\it Gaia} spectrum and wind model   }
	\label{fig:gaia_spec}
\end{figure}

\subsection{Neon}

Figure~\ref{fig:gaia_spec} shows the {\it Gaia} BP/RP spectrum. It covers a much larger wavelength range to the red. The sensitivity is less than that of our other spectra, and there are calibration issues as mentioned before. The top panel shows the wind model overplotted with the {\it Gaia} spectra. It shows that the wind feature at 7500--8000\AA\ is well reproduced. We use this `above atmosphere' spectrum to discuss the neon abundance.

Table \ref{tab:lines} lists a number of expected neon lines. The chosen neon abundance was based on lines in our optical spectra. The predicted  Ne\,{\sc x} line at 6894\,\AA\ is too strong but this coincides with a strong telluric feature that was removed from the ground-based spectra. 

The bottom  panel shows the {\it Gaia} spectrum with the wind model (red, dashed; shifted down for clarity). The lower red line shows the hot-gas model. The blue-dashed line shows the same model but with three times higher neon abundance. The neon lines now disagree with the observed spectrum. From this we deduce an upper limit of Ne/O\,$<0.15$ by number.

The wind spectrum is seen to overpredict the line at 8200\,\AA. In the wind model this is a Ne\,{\sc vii} complex, while the hot-gas model has an O\,{\sc viii} line at this location. The model that is shown has a high neon abundance. Again this points at a lower neon abundance for the remnant.

\subsection{Magnetic fields}\label{sec:mag}

The presence of strong magnetic fields has been proposed by \citet{gvaramadze2019}. This is  based on predictions of WD merger models which yield $B \sim 200$\,MG \citep{ji2013}, and on the expectation that the stellar wind is magnetically driven from a very rapidly rotating star. For the SN~1181 stellar remnant, \cite{kashiyama2019} carried out magneto-hydrodynamic simulations that argue for a WD remnant with a strong magnetic field (20 -- 50 MG) but still an order of magnitude lower than \citet{gvaramadze2019}.

Magnetic fields have been proposed for hot ($T>60$\,kK), hydrogen-poor (DO) white dwarfs with  O\,{\sc viii} lines and other ultra-high excitation lines (UHE) in absorption.  \citet{Reindl2019} attributes these  coronal lines to gas trapped in an equatorial, magnetically-confined, shock-heated magnetosphere. The confinement requires
\begin{equation}
    \eta ={B^2 R_\ast^2 \over \dot{M} v_w}  >1,
\end{equation}
where $B$ is the magnetic field strength, $R_*$ the stellar radius, $v_w$ the wind speed, and $\dot{M}$ the mass-loss rate. For this star, the mass-loss rate from the {\sc cmfgen} models require $B>0.1$\,MG for $\eta>1$, a modest magnetic field for such stars. 

The hydrogenic O\,{\sc viii} lines can provide strong constraints on magnetic field strength due to  Zeeman splitting of the line into $\pi$ and $\sigma$ components.  For moderate fields, the central $\pi$ component remains unchanged in wavelength, whilst the two $\sigma$ components shift to either side. The relative strength of the $\pi$ component depends on the magnetic field's orientation angle: it is absent when the magnetic field points at the observer and it is equal in intensity to the sum of the $\sigma$ lines if the magnetic field is perpendicular to the line of sight. Each of the three Zeeman components can be considered as a Gaussian of width derived from the temperature.  The observed line is the sum of these three Gaussians. For small shifts much less than the width of each line, this sum is itself a Gaussian widened by the shift. For larger shifts the line profile deviates.

The shift $\Delta \lambda$ between the two $\sigma$ lines can be parametrized as $\Delta \lambda = \alpha B$, where $\alpha$ is a constant for each ion and transition. The value for $\alpha$ is tabulated by \citet[][their Table~1]{2002PPCF...44.1229B}.  For transitions considered here, that is O\,{\sc viii} 8-9, O\,{\sc viii} 9-10, and C\,{\sc vi} 7-8, and for field strengths in Gauss, the $\alpha$ parameter is $1.8 \times 10^{-5}$, $3.4 \times 10^{-5}$, and $2.6 \times 10^{-5}$, respectively. The 6064\,\AA\ O\,{\sc viii} line has a FWHM of 93\,\AA. Assuming a separation between the $\sigma$ lines equal to the FWHM would have strongly distorted the shape, we constrain $B<2.7$\,MG. The 4340\,\AA\ line with FWHM of 69\,\AA\ gives a conservative upper limit of 3.8\,MG. 

We ran models adding the Zeeman components assuming all three Zeeman lines have equal intensity. This ratio corresponds to a magnetic field  orientation to the line of sight of 55 degrees. We then fit the profile of the 6064\,\AA\ line, where the intrinsic width is a fitting parameter. For $B=1.2$\,MG an acceptable fit to the line shape is found. For $B=1.5$\,MG the line profile deviation was clearly visible at 6064\,\AA, and developed a flat peak. Running the same test on the 4337\,\AA\ line shows it is indeed a bit less sensitive. We also ran the test with the central $\pi$ line twice as strong as each accompanying $\sigma$ line corresponding to a magnetic field oriented along the line of sight. Here the line remains centrally peaked which makes it easier to fit the profile. A notable deviation from the observed profile of the 6064\,\AA\ line was found for a field of 2.5\,MG. At this field strength, half the line width is due to Zeeman splitting. 

The O\,{\sc viii} line shapes give an upper magnetic field limit of $B<2.5$\,MG, easily within the requirement of the  magnetically-confined toroidal structure proposed by \citet{Reindl2019}. Based on the available data we can not rule out the alternative of such a structure.

This new field strength limit is far below that proposed by \citet[][200~MG]{gvaramadze2019} and of typical low-mass, magnetic WDs (10\,MG). There is some caution here. If the hot gas is located in the wind, then the surface magnetic field may be underestimated because the field strength of a magnetic dipole will go down as $1/r^2$ with distance $r$. Also, the star has a very high luminosity and temperature, and so a larger radius than a typical WD. If it evolves as a WD it will contract significantly, and the field would strengthen. Hence, our limits are not inconsistent with known field strengths of magnetic WDs, which are of order 10\,MG but they do not support values as high as 200\,MG previously proposed.

\begin{figure}
    \centering
    \includegraphics[width=\columnwidth]{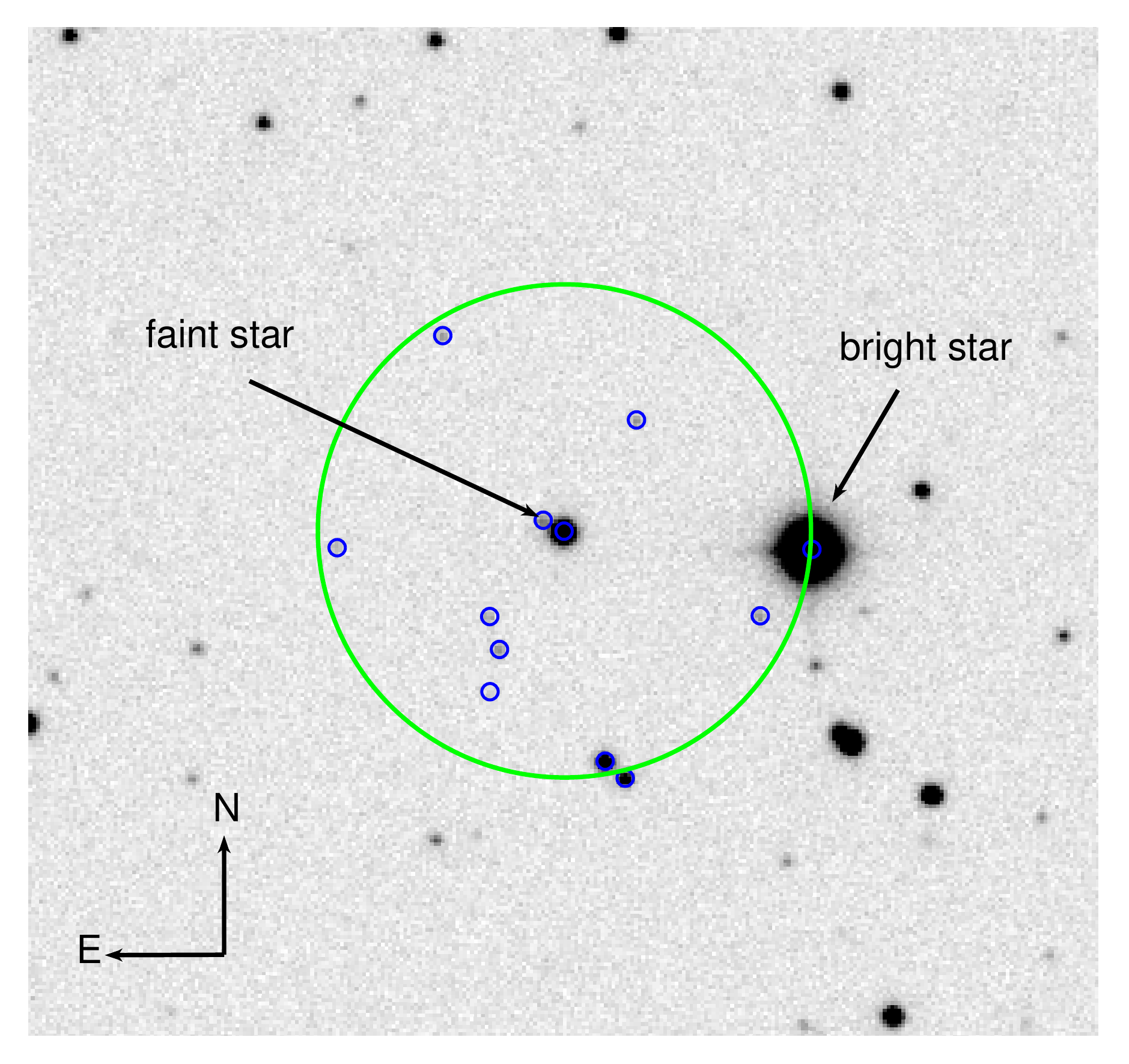}
    \caption{IPHAS $r$-band image of Pa\,30 identifying the faint and bright stars mentioned in Sect.~\ref{sec:varphot}. The green circle (radius 22\arcsec) is centered on the stellar remnant and it indicates the distance to the brightest star in the field. {\it Gaia} DR3 sources within a 25\arcsec\ radius from the remnant are marked as blue circles.}
    \label{fig:iphas}
\end{figure}

\section{Stellar variability}\label{sec:varphot}

\subsection{Photometry}

The stellar remnant ({\it Gaia} DR3 526287189172936320) has been observed by various terrestrial and space-based all-sky photometric surveys over many decades. Some offer short cadences to detect transient events (e.g., TESS 2-min and 30-min cadence), while others cover decades of sparse photometric sampling with long exposures (e.g, DASCH). Many have variable depth and sensitivity and large effective detector pixel sizes and large photometric apertures. This can result in blending and nearby source contamination.

A faint, foreground star ({\it Gaia} DR3 526287189166341504) is 2.3\arcsec\ eastwards of the stellar remnant (Figure~\ref{fig:iphas}). Though this star is clearly distinguished in short-exposure IPHAS \citep{Drew2005} images, it is $\sim4.5$ mag fainter in {\it Gaia} filters and well below limits of earlier photographic imagery so should not affect recorded optical photometry. A very bright star ({\it Gaia} DR3 526287257892412928) is  $\sim$21\arcsec\ directly west. Because recent all-sky surveys use large photometric apertures (e.g., the ASAS-SN PSF is $\sim$15\arcsec, or single-pixel aperture $\sim$17\arcsec\ in OMC), the bright star ($G\sim11$ mag) can affect adjacent pixels and influence photometric results. However, it is well separated from the stellar remnant in scanned data from all older photographic plates recorded in DASCH and so it is unlikely any variability for the stellar remnant is affected by the brighter star in 20$^{\rm th}$ century observations.

\begin{figure}
    \centering
    \includegraphics[width=\columnwidth]{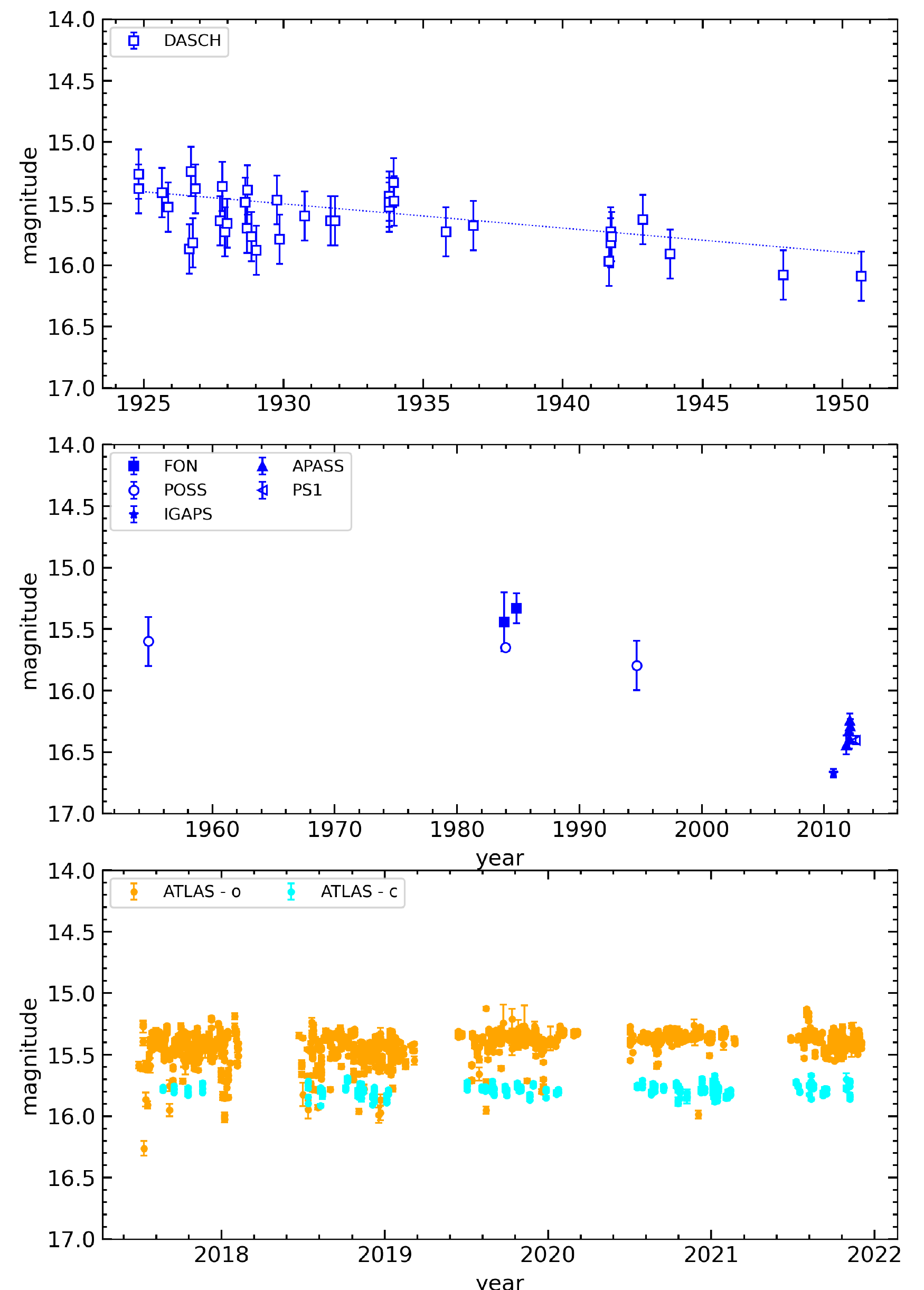}
    \caption{Extended light curve of the stellar remnant from 1920 until today. {\bf Top}: DASCH $B_j$ GSC calibrated photometry from 1920 to 1950 with the best-fit linear trend line plotted. {\bf Middle}: limited photographic (pre-2000) and CCD (post-2010) optical photometry in the $B$-band. Respective surveys are indicated in the legend. {\bf Bottom}: ATLAS binned photometry in the ``cyan'' ({\it c}) lower band (4200-6500\AA) and the ``orange'' ({\it o}) upper band (5600-8200\AA).}
    \label{fig:LC}
\end{figure}

\begin{figure}
    \centering
  \includegraphics[width=\columnwidth]{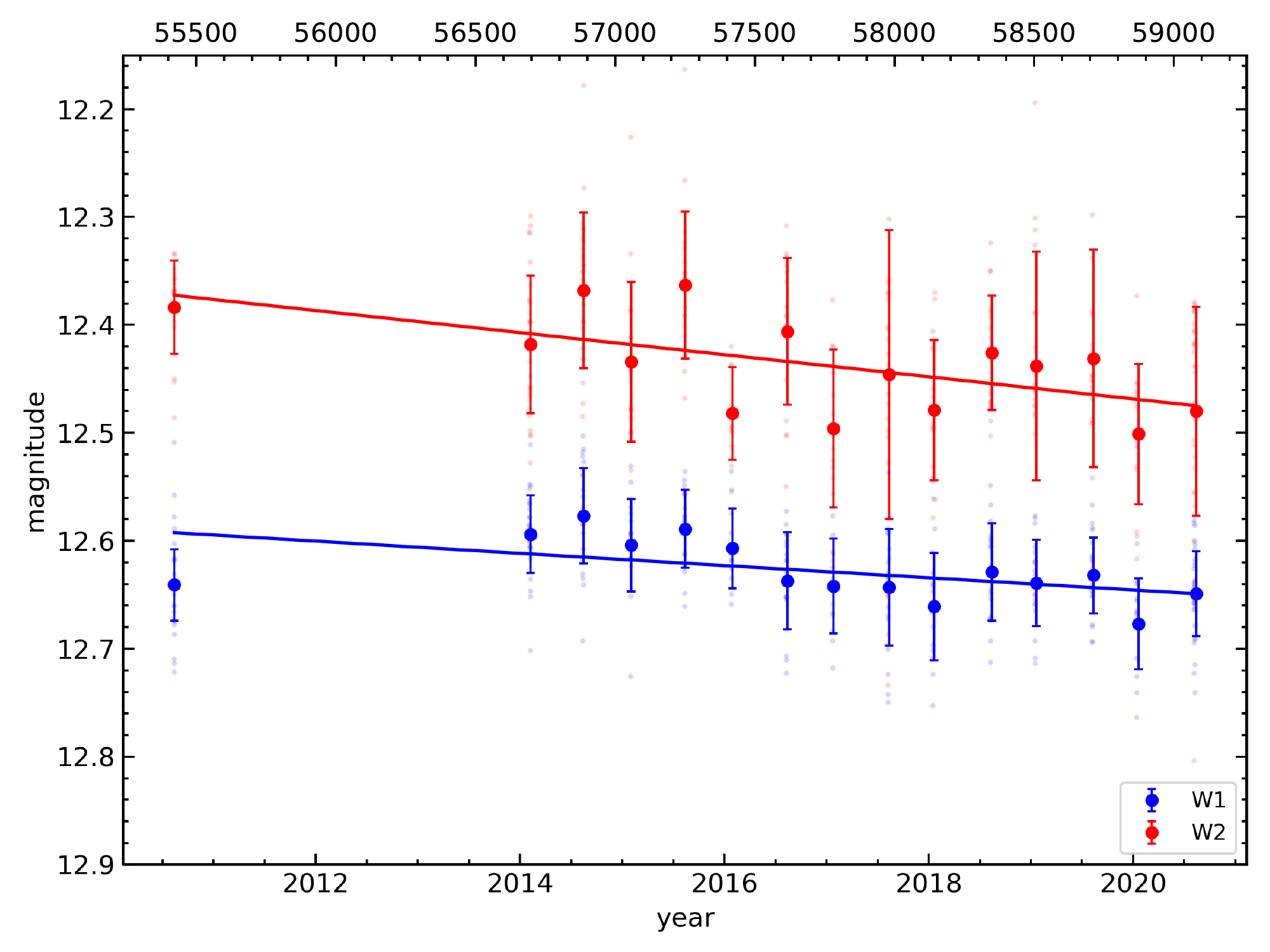}
    \caption{WISE $W1$ (blue; 3.4\micron) and $W2$ (red; 4.6\micron) light curves binned per epoch. The errors are the standard deviations of the binned measurements. The abscissa is given in years and Mean Julian Date.}    
    \label{fig:neo}
\end{figure}

Here, we show $B$ band results from various surveys as this filter has the best temporal coverage. The data include photographic photometry from DASCH, the Ukrainian FON astrographic catalogue \citep{2016KPCB...32..260A} and the Palomar Observatory Sky Surveys (POSS-I and POSS-II); CCD data from the AAVSO Photometric All-Sky Survey \citep[APASS DR10, ][]{2018AAS...23222306H}, the INT Galactic Plane Survey \citep[IGAPS/UVEX; $g,r$;][]{2020A&A...638A..18M}, the Panoramic Survey Telescope and Rapid Response System \citep[Pan-STARRS1, $g,r,i$;][]{tonry2012}, the Asteroid Terrestrial-impact Last Alert System survey \citep[ATLAS, $o,c$;][]{2018PASP..130f4505T}, the Transiting Exoplanet Survey Satellite \citep[TESS;][]{tess}, and the Near-Earth Object Wide-field Infrared Survey Explorer Reactivation Mission \citep[NEOWISE, $3.6, 4.5\rm\mu m$; ][]{2014ApJ...792...30M}. The single epoch IGAPS/UVEX $g,r$ photometry from Pan-STARRS1, and the Pan-STARRS1 $g,r,i$ photometry, were converted to $B$ using the transformations of \citet{tonry2012}. The optical light curves are shown in Figure~\ref{fig:LC} and the infrared light curve is shown in Figure~\ref{fig:neo}.

The stellar remnant is faint and the optical light curve sparsely populated over much of its recorded history (Figure~\ref{fig:LC}) but evidence of dimming is seen. The photographic $B_j$ magnitude estimates, calibrated via the GSC~2.3.2 \citep{2008AJ....136..735L} provided by DASCH for measurements between 1924.8 and 1950.7, show a mean $B_j$ magnitude of $15.6\pm0.2$ mag, and maximum magnitude variation of 0.85 from 37 data points with a dimming slope of 0.02~magnitudes/year and an $R^2=0.38$ (which indicates a weak though positive correlation). Exposures where the recorded plate limit was within 0.3~mags of the star's value are omitted as unreliable. Seven nearby stars, both fainter and brighter (including the bright star to the west), have much lower $R^2$ values (average of 0.047 with rms of 0.064) and are flat (average slope of 0.0007 mag/yr with rms also of 0.007). See the top panel in Figure~\ref{fig:LC}. Overall, the temporal coverage is rather sparse apart from the ATLAS data\footnote{Which has a pixel size of $\sim2$\arcsec\ in two bands: the ``cyan" ({\it c}) band (4200-6500\AA) and the ``orange" ({\it o}) band (5600-8200\AA).}. For the ATLAS dataset (before binning), no significant peaks were found in Lomb-Scargle periodograms so the source appears stable on short timescales in optical bands. 

We extracted archival photometry from the Wide-field Infrared Survey Explorer (WISE) 3-band Cryo data release \citep{2010AJ....140.1868W}, and the NEOWISE latest data release. Figure~\ref{fig:neo} shows the averaged photometry per epoch in bands $W1$ (3.4\micron) and $W2$ (4.6\micron). Each epoch covers about six months. Frame 
inspection showed there may be bright star contamination within the standard photometric aperture of NEOWISE (radius 8.25\arcsec). There is no nebula contamination as this is not visible at those wavelengths. The $W1$ and $W2$ light curves show a modest decline of $\sim$0.01~mag/year over 11 years, half the rate seen at earlier epochs in the optical $B$-band.

The wind model indicates some weak but no strong emission features in the WISE $W1$ and $W2$ filters: the emission is dominated by continuum. Since the Pa\,30 nebula is not visible at these wavelengths, we can exclude nebular lines such as [Fe {\sc ii}] and [Ar~{\sc vi}] or the 3.3\micron\ PAH band.  Therefore the  potential decline would be likely due to the continuum.

\begin{figure*}[htbp]
\gridline{\fig{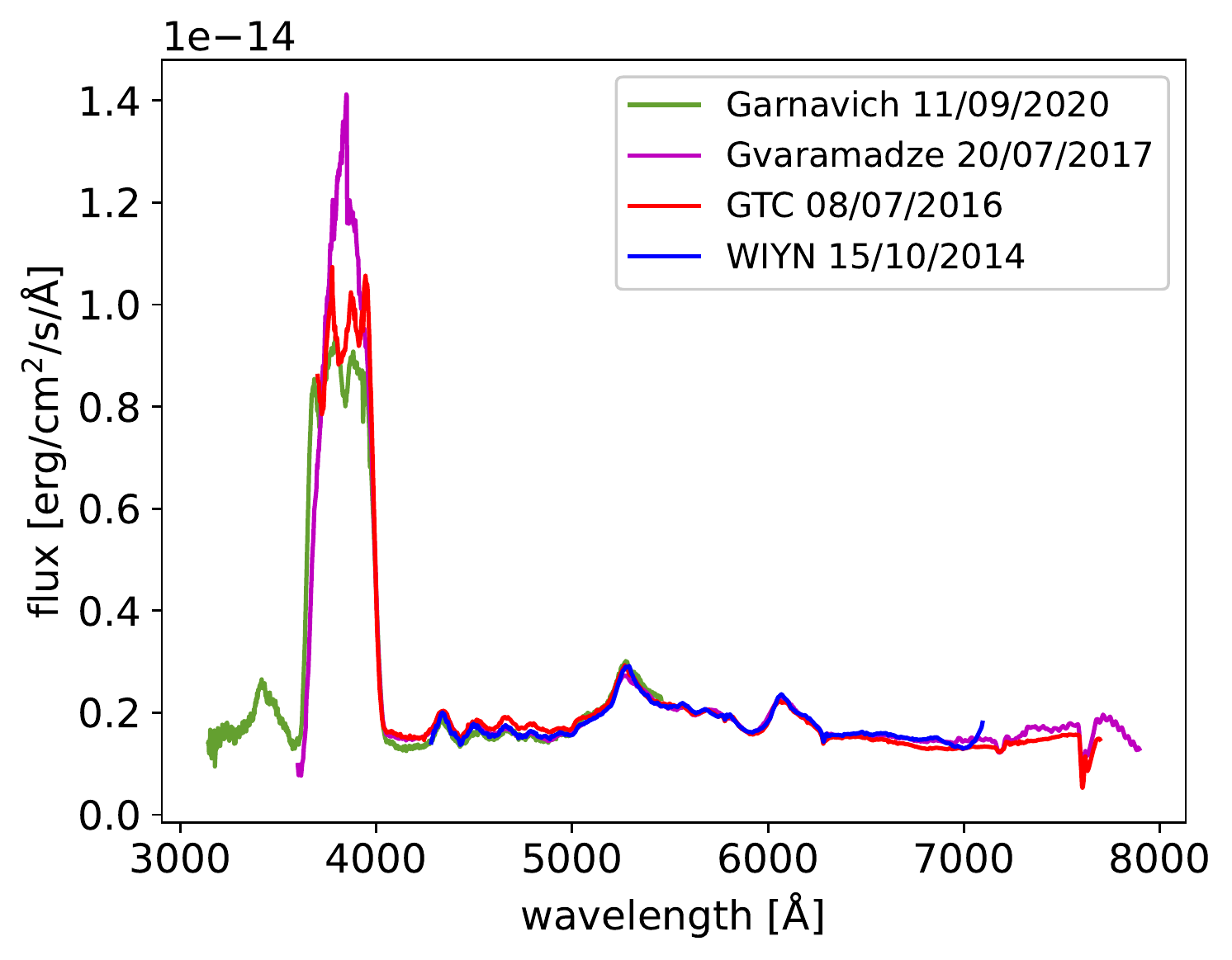}{0.3\textwidth}{(a)}
          \fig{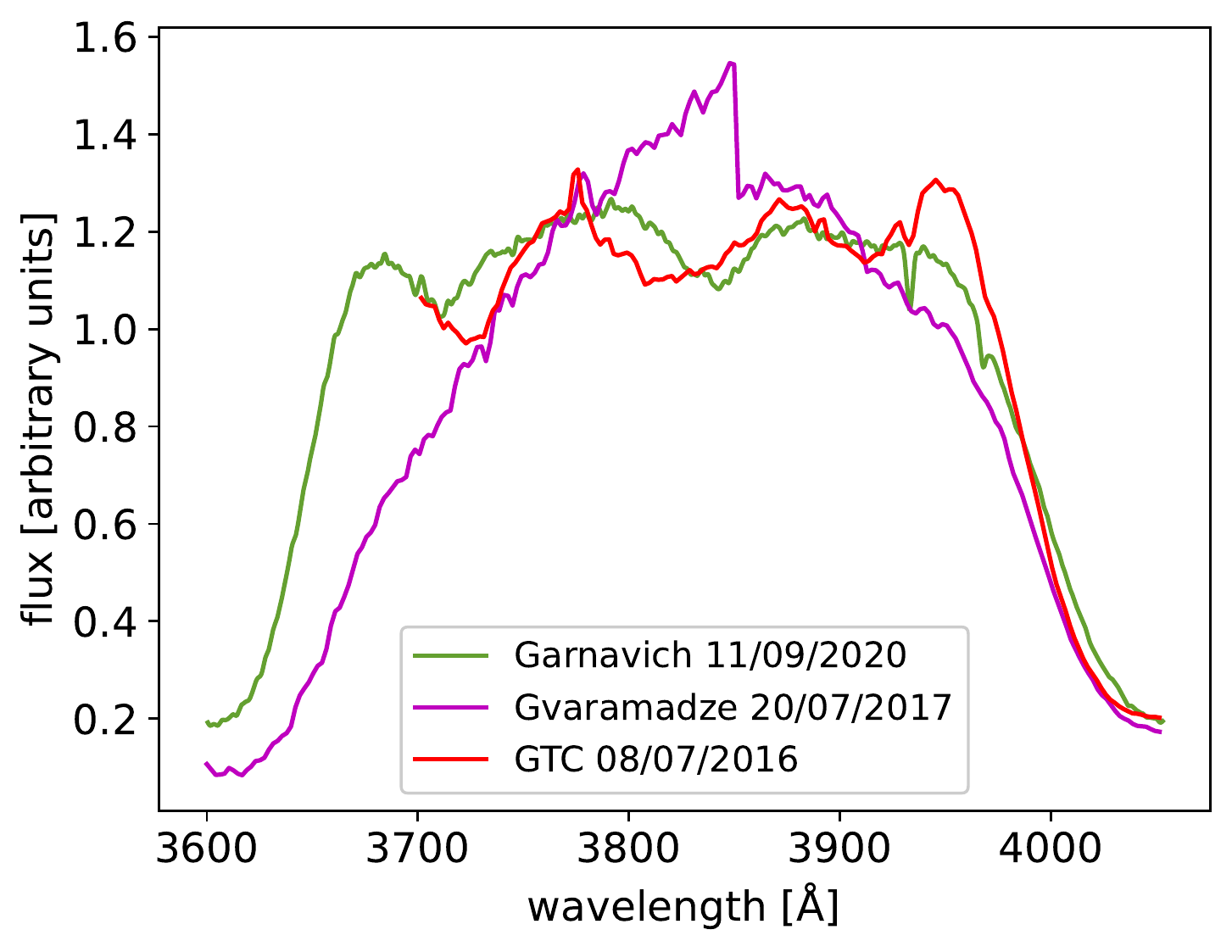}{0.3\textwidth}{(b)}
          \fig{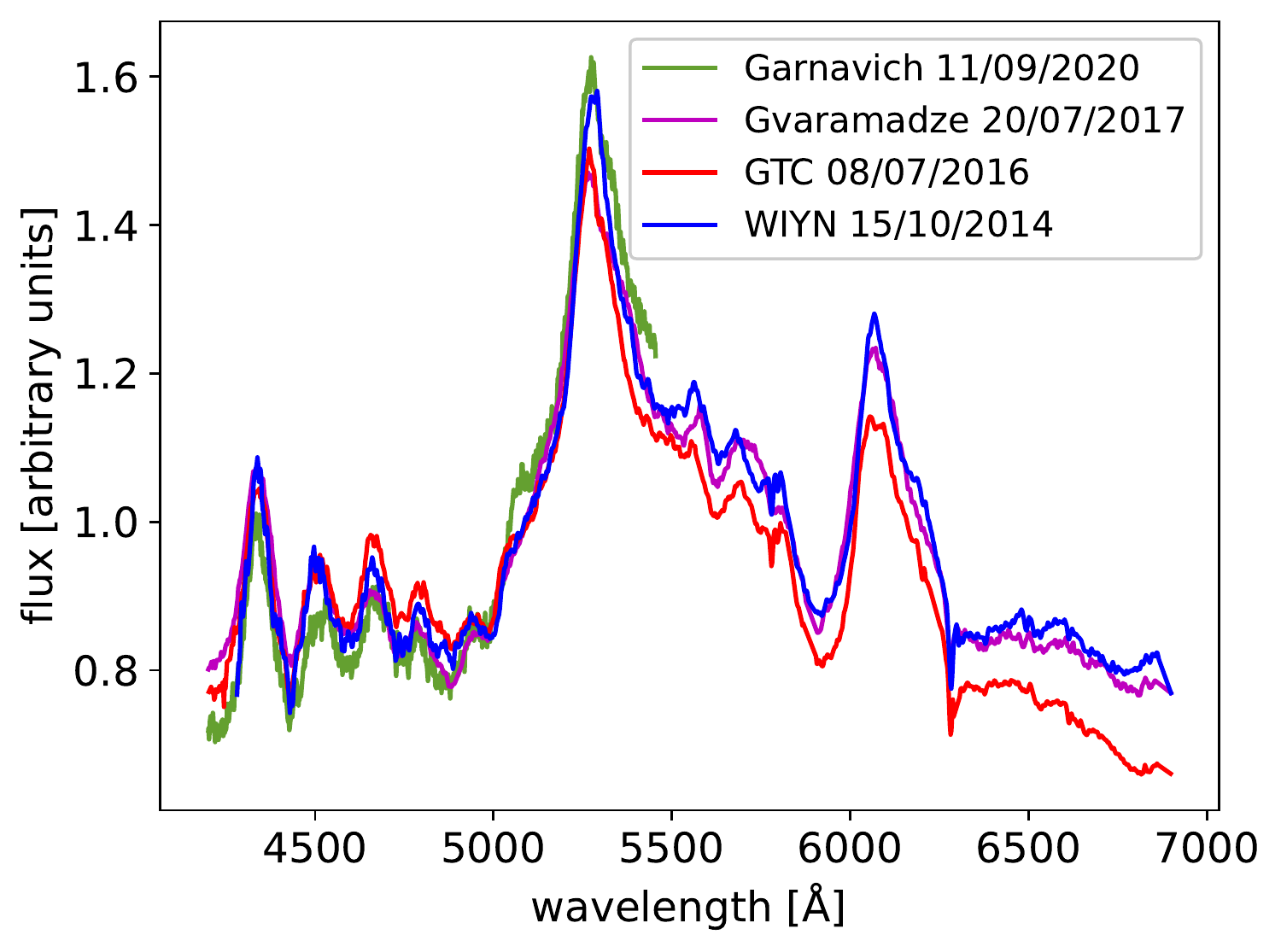}{0.3\textwidth}{(c)}
          }
\gridline{\fig{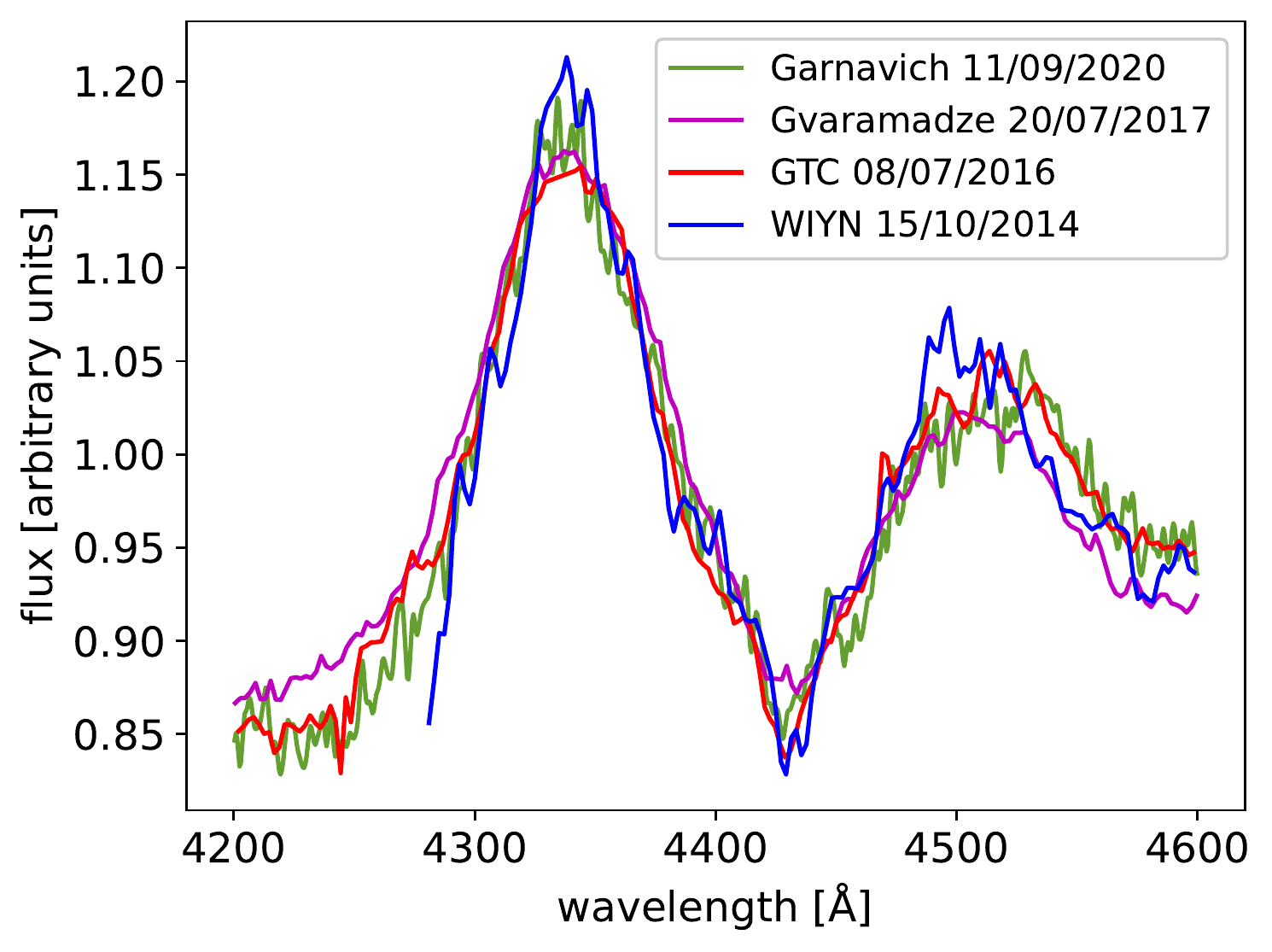}{0.3\textwidth}{(d)}
          \fig{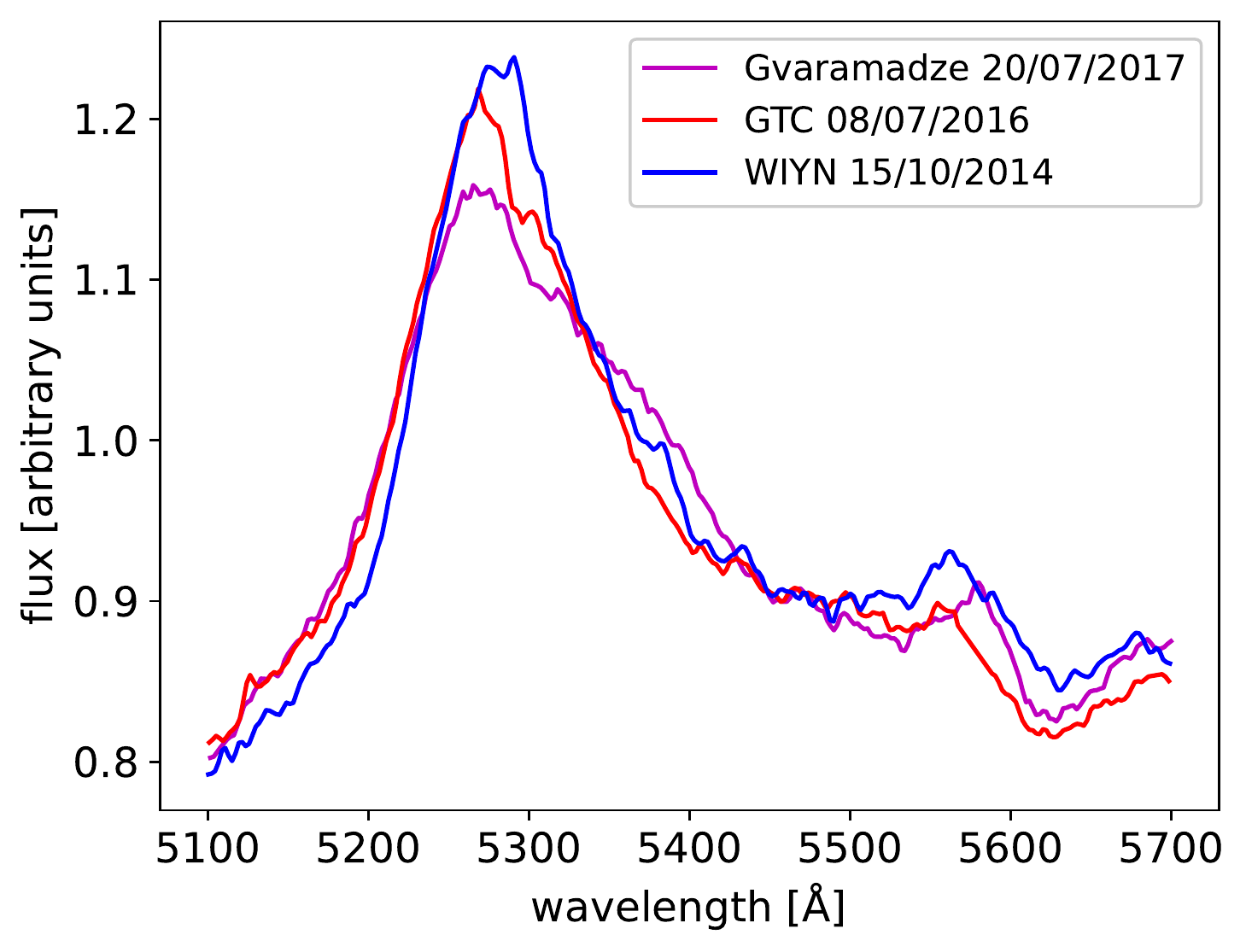}{0.3\textwidth}{(e)}
          \fig{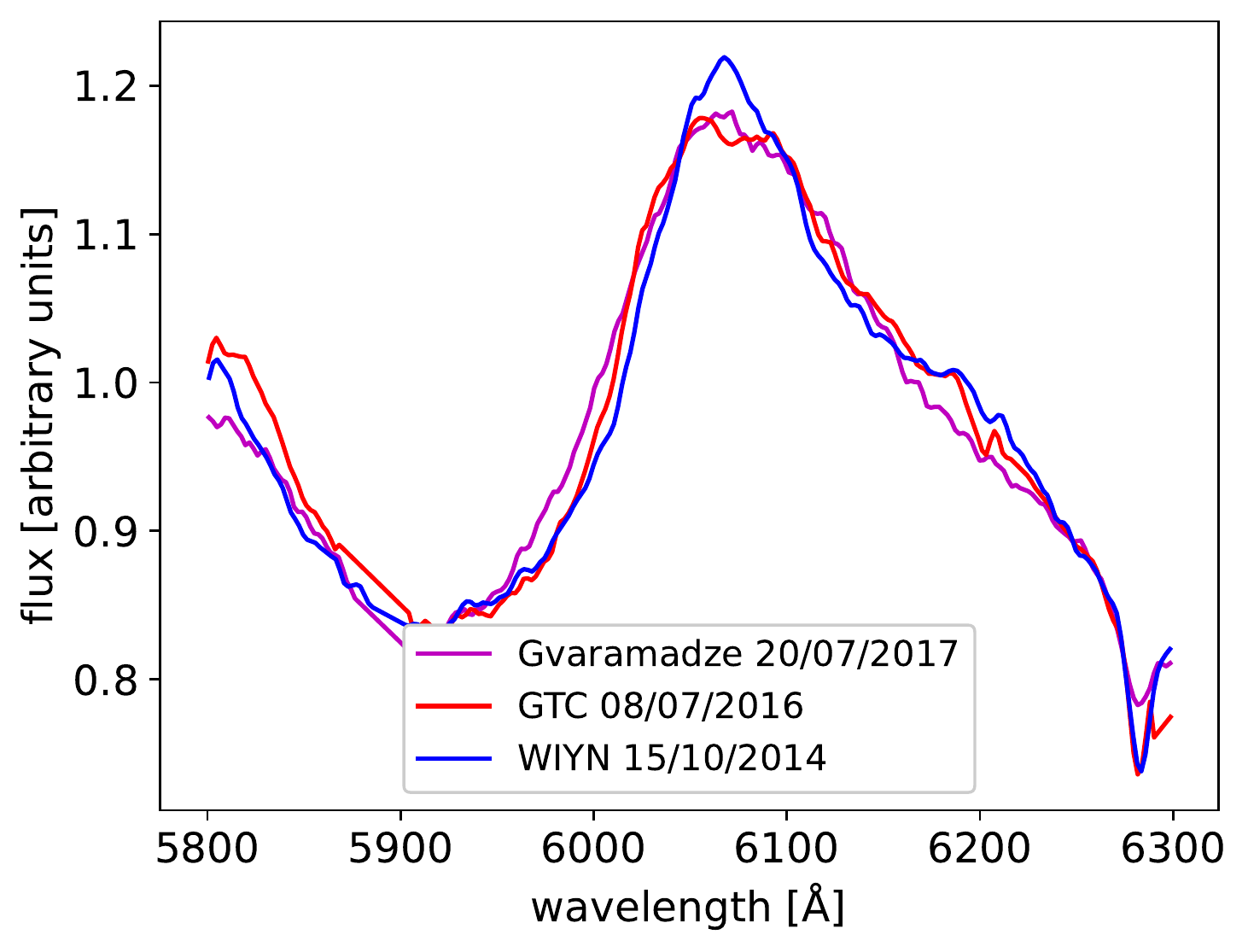}{0.3\textwidth}{(f)}
          }
\caption{Spectral differences between four epochs: our SparsePak/WIYN (2014) and OSIRIS/GTC (2016) against the \citet[][epoch: 2017]{gvaramadze2019} and \citet[][epoch: 2020]{2020RNAAS...4..167G}. In panel (a) all spectra are scaled to the SparsePak/WIYN (blue) levels to accommodate for slit losses. In all remaining panels spectra have been scaled to a mean value of 1.0 to achieve the best overlap and make differences in the shape of individual lines more obvious. See main text (Sect.~\ref{sec:varwind}) for more details.}
\label{fig:specdiff}
\end{figure*}

\citet{kashiyama2019} propose that the central star (WD J005311 in their paper) is a highly-magnetic WD spinning at an angular frequency of 0.2 to 0.5 s$^{-1}$. This would suggest very short light-curve variability at a period of 12.6 to 31.4 seconds. The current data does not trace this time scale.

\subsection{Wind variability}\label{sec:varwind}

In Section~\ref{sec:spec} we hinted at variability in the stellar wind. The SparsePak/WIYN data from 2014, our 2016 OSIRIS/GTC spectra, the 2017 spectrum by \citet{gvaramadze2019}, and the averaged spectra in 2020 by \citet{2020RNAAS...4..167G} are examined for long-term spectral variability (Figure~\ref{fig:specdiff}). 
The SparsePak/WIYN fibre is 4.7\arcsec\ and should include all stellar flux assuming the fibre is well placed. 
Comparison of the synthetic photometry from this spectrum (using the {\it Python} package {\it pyphot}\footnote{\url{https://github.com/mfouesneau/pyphot}}) to the photometric measurements provided by the individual surveys justifies this assumption. Hence, all slit spectra are scaled to match the SparsePak/WIYN spectrum. To improve the signal-to-noise ratio the SparsePak/WIYN and \citet{2020RNAAS...4..167G} spectra have been smoothed with a mean filter of length of 7 and 9 pixels respectively. 

The most striking differences are intensity changes in the O\,{\sc vi} 3811/34 emission line, as seen in panels (a) and (b) of Figure~\ref{fig:specdiff}. The discontinuity in the \citet{gvaramadze2019} spectrum at 3850\,\AA\ and the different spectral shape further to the blue are to be taken with caution, since flux calibration is less reliable in the far blue. \cite{2020RNAAS...4..167G} found that the O\,{\sc vi} 3811/34 line shows short-term $<2$hour variability, which was attributed to clumpiness in the stellar wind. Calibration differences among the different instruments do not allow a direct comparison of wind variability within the O\,{\sc vi} 3811/34 line, but the OSIRIS/GTC spectra overall shape appears consistent with the averaged spectra of \cite{2020RNAAS...4..167G}. A decrease in the peak emission of the O\,{\sc vi} 5300\,\AA\ line between 2014 and 2017 (panel (e) in Figure~\ref{fig:specdiff}) is also seen, perhaps associated with a change in the stellar wind. Small changes in continuum slope beyond 5500\,\AA\ can be due to instrumental and calibration effects between epochs, as the photometry appears to have remained constant throughout (cf. Sect.~\ref{sec:varphot}). Future photometric monitoring observations should be obtained at cadences of 10~seconds or less in the visual and at angular resolutions sufficient to allow separation of the target from the adjacent bright star (e.g., $\le 5$\arcsec). 

\begin{figure}[tbp]
	\centering
    \includegraphics[width=\columnwidth]{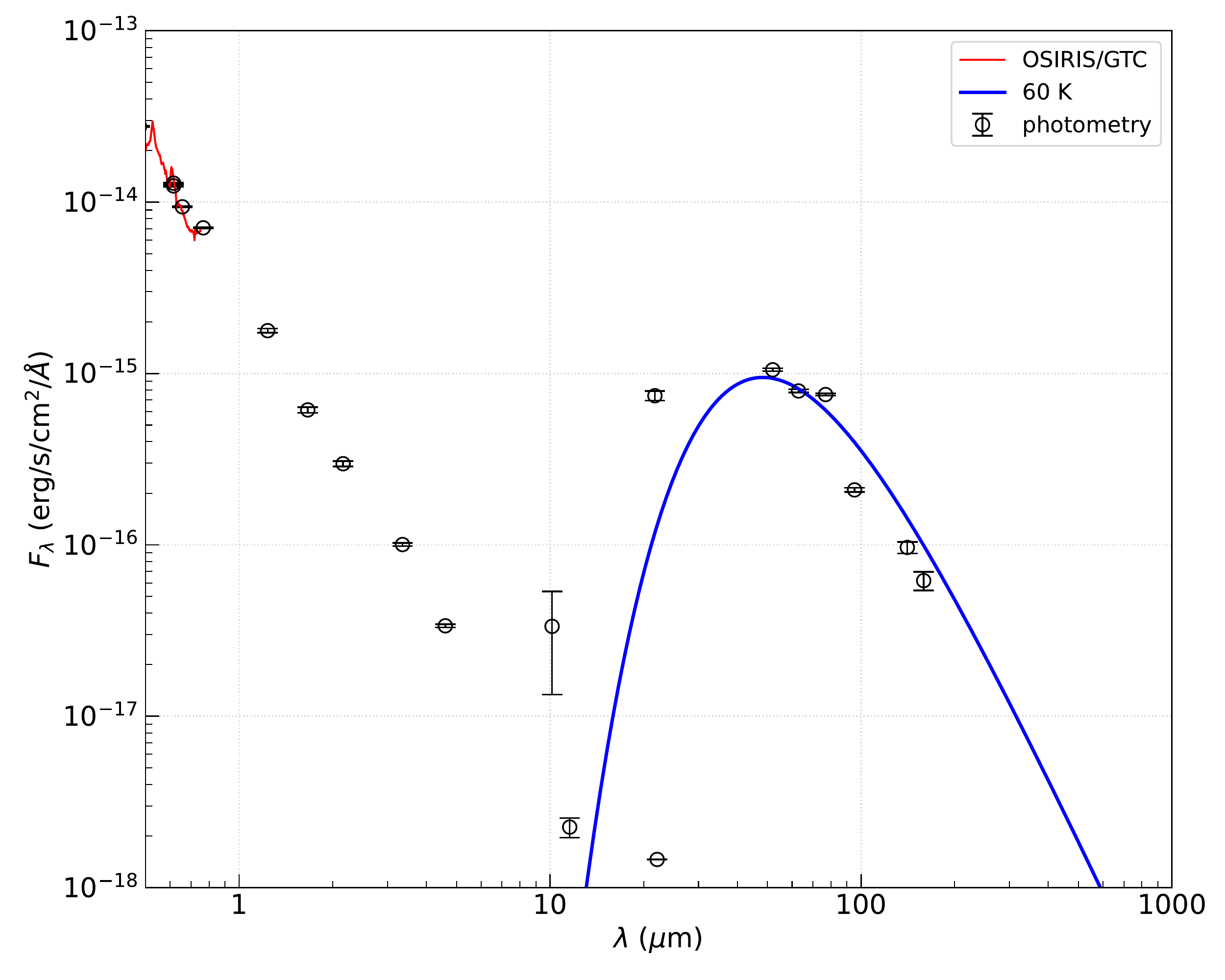}
	\caption{Infrared spectral energy distribution of Pa\,30 including the new photometry  in Sect.~\ref{sec:shellphotom} (60~K blackbody; blue) and the OSIRIS/GTC spectrum (red). There is no IR excess below 10\micron. The photometry below 3\micron\ and spectrum have been dereddened using extinction from \citetalias{paper1}.
		}
		\label{fig:SED}
\end{figure}

\section{The outer nebula} \label{sec:mass}

\subsection{Hydrogen emission}\label{sec:hydrogen}

Although no hydrogen lines were detected in the nebular spectrum from large aperture telescopes \citepalias{paper1}, the outer Pa\,30 nebula is tentatively detected in deep, $\approx$1~Rayleigh sensitivity\footnote{Compared to the $\simeq$3 Rayleighs for IPHAS where only the central star is visible.} H$\alpha$ imaging from the low angular resolution (pixel size 1.6\arcmin) Virginia Tech ``H$\alpha$" Spectral-line Survey \citep[VTSS;][]{vtss} that revealed a single enhanced pixel at Pa\,30's position. The 5-$\sigma$ excursion within a 15\arcmin$\times$15\arcmin\ region indicates a low-surface brightness hydrogen (H$\alpha$) shell of similar size to the [O\,{\sc iii}] shell found in \citetalias{paper1}. The VTSS survey limit corresponds to $5.661\times 10^{-18}$ erg\,s$^{-1}$\,cm$^{-2}$\,arcsec$^{-2}$ at H$\alpha$. The continuum-corrected surface brightness within a 1\,pc radius of Pa\,30 is $\approx 5.6$ Rayleigh  $\approx 8.8\times10^{-13}$ erg\,s$^{-1}$\,cm$^{-2}$. Adopting the extinction law of \citet{howarth1983} for H$\alpha$, $c_{H\alpha}=0.99\,E(B-V)$, then the dereddened H$\alpha$ brightness becomes $7.4\times 10^{-12}$\,erg\,s$^{-1}$\,cm$^{-2}$ for $A_V=2.9$, which translates to $L_{\rm H\alpha}\leq1.2\,\rm L_{\odot}$. Deeper narrow-band H$\alpha$ imagery is needed to confirm this tentative detection.


\subsection{Infrared emission}\label{sec:shellphotom}

Pa\,30 is visible at mid- and far-infrared wavelengths though not detected in the IRAS 12\micron\ maps. In the far-infrared, the shell is round as indicated by AKARI 65 to 140\micron\ maps, extending from 88\arcsec\ up to 128\arcsec ($\pm5$\arcsec) from the central star, or else a radius of about 0.98 -- 1.42 pc at the adopted distance. Foreground interstellar emission confuses the images beyond 100\micron. We attribute the far-infrared wavelengths to dust emission and the mid-infrared emission to line emission. 

Given the infrared shell sizes and the uncertainties in archival point-source photometry, we have re-estimated surface brightnesses for all IRAS and AKARI images. Images were downloaded from IRSA/IPAC with the same angular resolution (15\arcsec/pixel): HIRES maps for IRAS, and Far-Infrared Surveyor maps for AKARI. An 100\arcsec\ radius aperture was selected for all maps. We use the conversion factor of \citet{ueta2019} to derive flux densities. The final photometry beyond 100\micron\ was corrected for interstellar contamination (Table~\ref{tab:dmass}). IRAS 12\micron\ and AKARI 160\micron\ flux densities are upper limits (non detection). The new measurements are given in Figure~\ref{fig:SED}, along with archival photometry from Pan-STARRS1, {\it Gaia}, IPHAS, 2MASS, and WISE. The IRAS and AKARI fluxes integrated over the nebula indicate the shell flux peaks at $\lambda \sim$90\micron, while the foreground emission is colder. Using the aperture photometry results for the nebula and the VOSA tool, we derive a shell blackbody temperature of 60~K (cf. Figure~\ref{fig:SED}). 

\begin{table}
	\centering
	\caption{Pa~30 nebula surface photometry vs. {\sc cloudy} predictions.}\label{tab:dmass}
	\begin{tabular}{ccc}
	\hline
	Band ($\mu$m) & $F_{\nu}$ (Jy) & {\sc cloudy} model (Jy) \\ \hline
	IRAS/12$^\dagger$ & $\leq0.25$ & 0.25  \\
	IRAS/25 & $1.41\,(\pm0.09)$  & 1.4 \\
	IRAS/60 & $13.24\,(\pm0.26)$ & 16.6 \\
	AKARI/65 & $11.72\,(\pm0.25)$ & -\\
	AKARI/90 & $19.99\,(\pm0.32)$ & -\\
	IRAS/100 & $7.25\,(\pm0.20)$  & 7.8 \\
	AKARI/140 & $\leq6.39\,(\pm0.51)$ & -\\
	AKARI/160$^\dagger$ & $\leq5.27\,(\pm0.66)$ & 2.0 \\
	\hline
	\end{tabular}
	\tablecomments{($\dagger$) stands for upper limit. }
\end{table}

We ran a {\sc cloudy} model for the shell assuming it is ionized gas with normal ISM abundances. We assume a blackbody star with $L_* =3\times 10^4\,\rm L_\odot$, $T_* = 2.3 \times 10^5\, \rm K$, surrounded by a shell with  inner and outer radius of 0.8 and 1.2\,pc, and density $n= 17\,\rm cm^{-3}$. Standard ISM abundances and dust content were used as defined in {\sc cloudy}, with a range of grain size and a dust-to-gas ratio of $3.8\times 10^{-3}$. This gives a total gas mass of $\sim 2.1\,\rm M_\odot$ and a dust mass of $8 \times 10^{-3} \,\rm M_\odot$. The model predicts broadband fluxes listed in Table \ref{tab:dmass}, in good agreement with the infrared measurements. At shorter wavelengths, the model flux is dominated by a few strong emission lines: [Ne\,{\sc vi}] at 7.6\micron, [Ne\,{\sc v}] at 14.7\micron\ and 25\micron, and [O\,{\sc iv}] at 25.9\micron. The longer wavelength emission cannot be explained by emission lines and is from dust. The predicted H$\alpha$ flux is $2.2\times 10^{-11}\,\rm erg\,s^{-1}\,cm^{-2}$, which, after extinction correction, gives 13 Rayleighs, 2.3$\times$  the VTSS H$\alpha$ surface brightness. The model explains the infrared photometry but overpredicts the apparent H$\alpha$ surface brightness.

Alternatively, it can be assumed that the H$\alpha$ emission comes from the accelerated gas seen in the [S\,{\sc ii}] emission reported in \citetalias{paper1}. The average density of this component, derived from the [S\,{\sc ii}] lines, is $N_{\rm e} = 120\,\rm cm^{-3}$. We assume a temperature $T_{\rm e} = 10^4$\,K and use the relation \citep{pottasch1984}:
\begin{equation}
M_{\rm ion}\,[{\rm M_\odot}] = 11.06 \times F({\rm H}\beta) \, \frac{d^2 \,(T_{\rm e}/10^4)^{0.88}}{N_{\rm e}},
\end{equation}
where $F({\rm H}\beta)$ is in units of $10^{-11}$ erg\,cm$^{-2}$\,s$^{-1}$, distance $d$ in kpc, electron temperature $T_e$ in K, and electron density $N_e$ in cm$^{-3}$. The intrinsic H$\beta$ flux, derived from the H$\alpha$ flux assuming recombination Case B for a theoretical H$\alpha$ to H$\beta$ ratio of 2.85, is $F(\rm H \beta)\approx 7.2 \times10^{-12}$ erg~cm$^{-2}$~s$^{-1}$. Hence, an ionized mass of $\approx0.3\,\rm M_\odot$ is derived.

The mass is therefore dominated by the outer shell, rather the accelerated gas seen in [S\,{\sc ii}]. The overprediction of H$\alpha$ for the outer shell may indicate that this gas is hydrogen-poor, in which case the mass is less than the derived 2.1\,M$_\odot$.

The presence of dust in the circumstellar shell is strongly supported.  The mass is dependent on composition and grain size. The dust origin remains open: it can be supernova dust, interstellar dust heated by the star, or dust from previous mass loss. The circumstellar shell could represent swept-up ISM or matter ejected in a mass-loss episode before the supernova explosion, such as a relic planetary nebula. 

\section{Discussion}\label{sec:discussion}

\subsection{SN shell mass}

\citet{oskinova2020} estimate the amount of hydrogen-free, X-ray-emitting nebula gas to be $0.1~\rm M_\odot$. Our ionized mass estimate from the tentative H$\alpha$ detection is an additional 0.3 to 2.1 $\rm M_\odot$ (Sec.~\ref{sec:shellphotom}). However, the relation of this gas to the supernova is unclear (see next subsection) and its kinematics are not known.

Consider the kinetic energy of the H-poor ejecta, $E_{\rm kin} = (1/2)\, M_{\rm ej} v^2_{\rm exp}$,
where $v_{\rm exp}$ is the ejecta expansion velocity. In \citetalias{paper1}, we find the shocked emission has a radial velocity of 1100\kms. If the extended halo-like feature in the AKARI 90\micron\ data (radius $128\pm5$ arcsec) is the farthest extent of the nebula, then the ejecta velocity would be $\sim$1700\kms\, not much different to the shocked emission. Both values are consistent with the general decline of ejecta velocities more than 100 days post-eruption in Type Iax SNe \citep[e.g.,][]{2018PASJ...70..111K,2021arXiv211109491S}. 

Using the above ejecta mass and velocities, the kinetic energy of the ejecta $E_{\rm kin}$ ranges from $1\times10^{48}$ to $3\times10^{48}$ erg. These are conservative estimates since this is a lower limit for ejected mass, while the expansion velocity at the time of eruption could have been higher.

\begin{figure}
    \centering
    \includegraphics[width=\columnwidth]{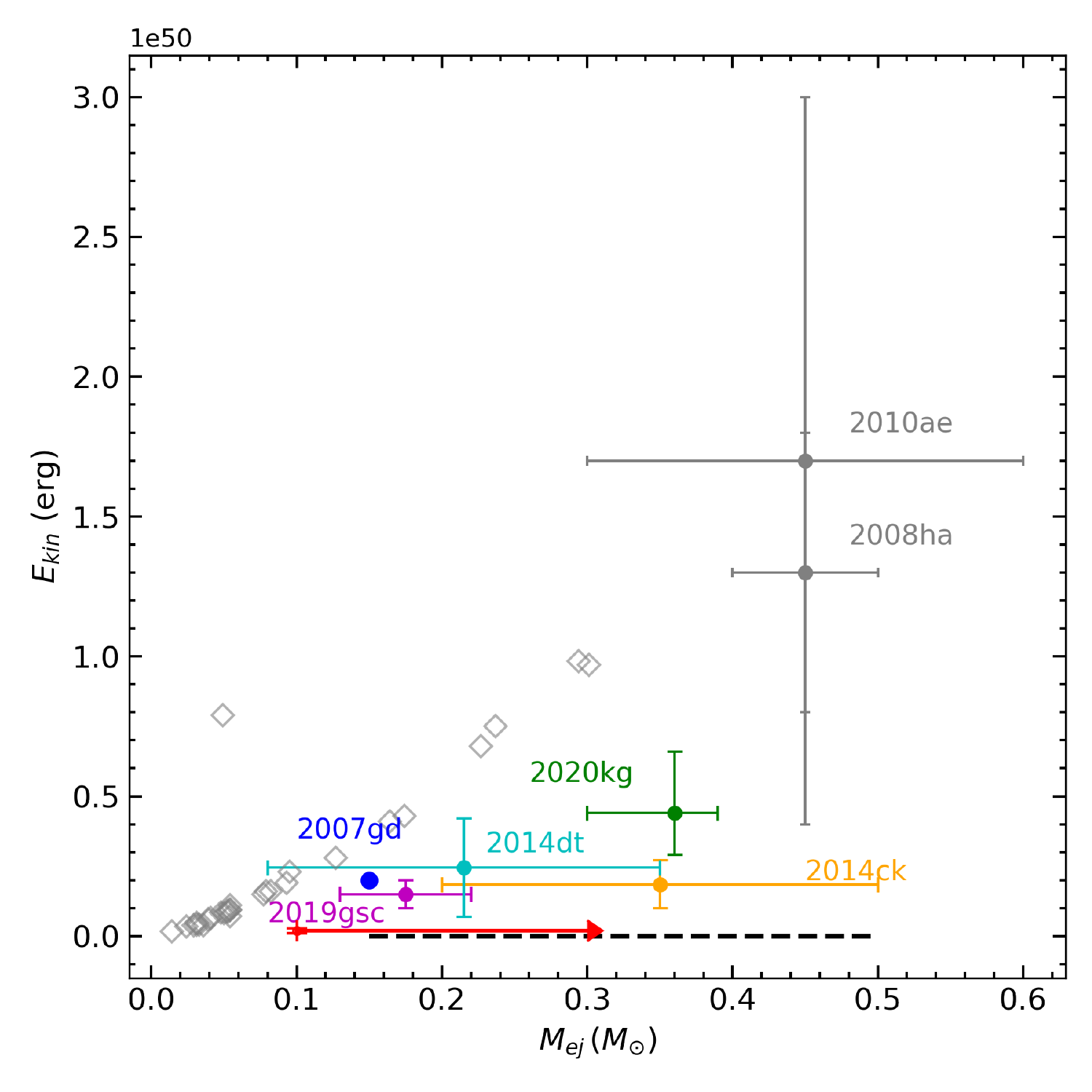}
    \caption{Comparison of the ejecta mass against kinetic energy released by Type~Iax SNe. Pa~30/SN~1181 is marked in red. \citet{lach2022} models are also shown (open diamonds), while the range of expected values for Iax by \citet{2013ApJ...767...57F} is shown in black (dashed). A few examples of known extragalactic Iax are also shown: 2008ha and 2010ae data by \cite{2014A&A...561A.146S}, 2007gd by \cite{2010ApJ...720..704M}, 2014ck by \citet{2016MNRAS.459.1018T}, 2019gsc by \citet{2020ApJ...892L..24S}, and 2020kg by \cite{2021arXiv211109491S}.
    }
    \label{fig:mej}
\end{figure}

We compare these tentative estimates against known extragalactic Type~Iax SNe in Figure~\ref{fig:mej}. These include theoretical models by \citet[][open diamonds]{lach2022}, the proposed range of values for Type~Iax by \citet[][black dashed bar]{2013ApJ...767...57F}, and a few examples of extragalactic sources: 2008ha and 2010ae \citep[gray; ][]{2014A&A...561A.146S}, 2007gd  \citep[blue;][]{2010ApJ...720..704M}, 2014ck \citep[orange;][]{2016MNRAS.459.1018T}, 2019gsc \citep[magenta;][]{2020ApJ...892L..24S}, and 2020kg \citep[green;][]{2021arXiv211109491S}. Both the low $E_{\rm kin}$ and ejecta mass estimate are consistent with observed ranges for Type~Iax sources. However, they do not agree with the model predictions for pure deflagrations of Chandrasekhar-mass CO WDs of \citet{lach2022}.

\subsection{Merger and ejected mass}

 The stellar luminosity and temperature place the star on a location in the HR diagram similar to post-AGB stars. In these stars, the stellar luminosity comes from residual hydrogen burning and follows the core mass--luminosity relation, originally proposed by \citet{Paczynski1970}. Using the star's luminosity determined as $L_* = (40\pm 10)\times 10^3\,\rm L_\odot$, this relation gives a core mass (which is essentially the mass of the remnant star) of $M_* =1.20\pm0.17\,\rm M_\odot$. The relation from \citet{VW1994} gives 0.1\,M$_\odot$ less.

 The application of these models to post-merger evolution can be queried. In the absence of detectable hydrogen, helium burning may be expected. The models of \citet{VW1994} find a similar luminosity for stable helium burning, however this helium very quickly runs out in post-AGB stars and the luminosity subsequently declines.

Merger models published by \citet{schwab2016} and \citet{schwab2021a} reach a similar location in the HR diagram as carbon burning stars, but take 10 to 20 kyr to reach this. They reach the post-AGB phases at 30\%\ lower luminosity than predicted by the \citet{VW1994} for their mass. Lower mass merger models by \citet{Schwab2021b} take 170 kyr to reach the post-AGB phase and similar models from \citet{Wu2022} reach the right luminosity but do so as cool red giant stars. All these merger models spend a long time as red giants. The models assume a red giant mass loss recipe. The evolution could conceivably be much accelerated if an instantaneous mass loss is assumed, either due to the explosion or akin to common envelope systems, thereby skipping the red giant phase and the early post-AGB phase. Accretion may also affect the evolution in the post-AGB region of the HR diagram. 

Assuming the change in luminosity by 30\%, the mass of the star would become $1.45\pm0.2\,\rm M_\odot$.  The mass range does not require the star to have super-Chandrasekhar mass as proposed by \cite{gvaramadze2019}, although this is not ruled out either.

The range of possible masses fits with WD merger products.  The CO WD mass distribution peaks at approximately 0.6\,M$_\odot$, while helium WDs are approximately 0.45\,M$_\odot$. Two merging CO WDs or a CO+He WD can therefore reach the minimum remnant mass. O-Ne WDs are more massive and a merger involving one \citep{gvaramadze2019} would lead to a more massive remnant. The moderate neon abundance may not support involvement of such a WD. 

The mass ejected in the merger should be included in the progenitor masses. The minimum ejecta mass is that derived from the X-ray emission \citep{oskinova2020} of 0.1\,M$_\odot$. The outer shell may contain further mass ejected during the merger. The hydrogen mass in the shell cannot come from the merger as WDs have very thin hydrogen layers and this should not be be counted in the ejecta mass. The hydrogen is seen only in the VTSS detection and needs confirmation. If the outer shell is hydrogen-free, then it may add another  0.1\,M$_\odot$. The merger ejecta are therefore estimated at $M_{\rm ej} = 0.15 \pm 0.05\,\rm M_\odot$.  

This leaves the question of where the outer shell originates. If there is hydrogen in the outer shell, it can be swept-up ISM or from mass lost by the WD progenitor. If so the faint, outer shell is a relic planetary nebula, re-ionized by the luminous, hot post-merger star.

\subsection{Dust}

Pa\,30 is only the second detection of dust emission in Type Iax SNe and the only known case of cold dust.  The other example is the recent Type Iax SN~2014dt in M~61. {\it Spitzer} observations by \citet{2016ApJ...816L..13F} found excess mid-infrared emission (3.6 and 4.5$\rm\mu m$) one year post-eruption in 2014dt, that could be a pre-existing dusty circumstellar matter or newly formed dust ($\sim10^{-5}\,\rm M_\odot$ of carbonaceous material). Coincidentally, 2014dt has been suggested as a possibly bound remnant \citep{2018PASJ...70..111K}. We note that cold dust, as in Pa\,30, would not be detectable in extragalactic SNe.

\subsection{The lack of a kicked stellar remnant}

Theoretical models of Type~Iax explosions predict a kicked stellar remnant \citep[e.g.,][]{jordan2012,kashyap2018}. However, the radial velocity of Pa\,30's stellar remnant, as well as the mean radial velocity of its surrounding nebula, is nearly zero, while the tangential velocity is low ($\sim30$\kms). The {\it Gaia} DR3 proper motion indicates the central star could have moved by at most $\sim2$\arcsec\ since 1181~AD. It clearly remains at the center of Pa\,30. So these data are already a challenge to existing models. The nearby faint star (Figure~\ref{fig:iphas}) cannot be a kicked companion as this star is in the foreground \citep[$1.1^{+0.4}_{-0.3}$~kpc;][]{bailerjones2021} with a much higher proper motion (21.285 mas yr$^{-1}$)

\citet{lach2022} estimate kick velocities between 6.9\kms\ and 369.8\kms\ for bound remnants in Type Iax explosions. These could be in the ballpark for Pa\,30.
 
The (only) CO+ONe WD merger model is by \citet{kashyap2018} with a predicted kick velocity of $\sim 90$\kms. However, an ONe WD may not be required here because of the remnant mass and the moderate neon abundance.

\section{Conclusions}\label{sec:conclusions}

The association of Pa\,30 with the supernova of 1181~AD \citepalias{paper1} makes this system one of only two Galactic Type~Iax supernovae known, the other being Sgr~A East \citep{2021ApJ...908...31Z}. Pa\,30 is  the only bound SN with a visible stellar remnant in the Galaxy. Its Wolf-Rayet-type optical spectrum is the only example known that is neither the central star of a planetary nebula nor the product of a high-mass Population I progenitor. This system, unique in several respects, is the nearest and youngest Type~Iax remnant known and amenable to detailed study. It is also the only example where a Type~Iax stellar and nebular remnant can be related to an observed eruption almost a millennium earlier.

Most features of the optical spectrum can be reproduced with a stellar wind model, with $v_{\rm w} = 15,000\,\rm km\,s^{-1}$ and mass loss rate $\dot M \sim 10^{-6}\,\rm M_\odot\,yr^{-1}$. The wind appears H-free, is strongly deficient in helium relative to C, N, O, and Ne, and has C/O$\sim 0.25$ by number. The wind cannot be driven by radiation pressure alone.  Some broad O\,{\sc viii} lines not reproduced by the wind model can be well fitted with hot gas with $T\simeq 4$\,MK, either shocked gas in the outer or inner wind where the opacity of the wind favours the former. This gas, with the same abundances as the wind, is seen in lines of O\,{\sc viii} and possibly C\,{\sc vi}. The hot gas shows a low neon abundance, Ne/O$\,<0.15$ by number. 

The infrared spectrum can be fitted with a circumstellar shell where shorter wavelengths (10--30\micron) show line emission and longer wavelengths show emission from cold dust, with $M_{\rm dust} \sim 8\,\times\, 10^{-3}\,\rm M_\odot$.  The shell is also tentatively detected in wide-field H$\alpha$ imaging. This is only the second detection of dust in a Type Iax SN and the only detection of cold dust. The origin of this dust is open: possibilities are supernova dust, heated interstellar dust, or dust from previous mass loss from the system.

On timescales of the order of a century, the visual light curves of the central star show evidence of $\sim0.5$ magnitude dimming between 1925 and 1950 around a value of $B_j\sim$15.8 mag. Recent mid-infrared data from WISE may also show a slight fading over 11 years of about 0.1~magnitudes for $W1$ and $W2$.

The abundances, ejecta mass, explosion luminosity and energetics are all consistent with a type Iax supernova. There are two scenarios for this type of supernova: a pure deflagration of a Chandrasekhar-mass CO WD, and a double-degenerate merger. The observed parameters and the presence of a highly evolved post-eruption star indicate that the stellar remnant is an example of the second scenario: a merger of two WDs \citep[also proposed by ][]{gvaramadze2019,oskinova2020}.  The luminosity , as derived from the stellar wind modeling $L=(40\pm10)\times 10^3\,\rm L_{\odot}$ for the assumed distance and reddening,  indicates a mass of the remnant star of $1.2\pm0.2$\,M$_\odot$ to $1.45\pm0.2$\,M$_\odot$ dependent on the stellar models used for the post-AGB region  of the HR diagram.  This allows for the possiblity of masses below the Chandrasekhar mass, in which case  the object might not go supernova again \citep{gvaramadze2019}. The mass is consistent with a merger between two CO WDs or a high-mass CO+He WD. The low neon abundance may argue for the former. The ejecta indicate that $0.15\pm0.05$\,M$_\odot$ was lost during the merger. The outer shell, if indeed H-rich, cannot be purely explained as merger ejecta and could be swept-up ISM or a fossil planetary nebula from pre-merger mass-loss.

Merger products are  expected to be fast rotators. Our observations cannot establish a short rotational period and this remains to be tested. Merger products are also assumed to be the progenitors of highly magnetic WDs. The absence of detectable Zeeman splitting places an upper limit of $B<2.5$~MG, way below the previously predicted values from \citet[][200~MG]{gvaramadze2019} but may increase during future remnant contraction.

The mass of the merger progenitor stars, the ejecta and circumstellar mass, the non-detection of a significant magnetic field, the fact that the remnant has remained at the center of its bound nebula post-eruption, and the possibility of modest photometric dimming of the stellar remnant, are some of its key observed properties that that somewhat deviate from the typical (inferred or assumed) properties of Type Iax SNe, and therefore merit detailed investigation  to improve our understanding of Type~Iax evolutionary models.

\section*{Acknowledgements}

We especially thank the anonymous referee for their constructive review that helped improve our manuscript, as well as Dr.~G.~Gr\"afener and Dr.~P.~Garnavich for providing their spectra. We thank Dr.~Andr\'as P\'al for help with qualitative analysis of the TESS data. We thank Dr.~Laurence Sabin for designing the OSIRIS/GTC observations.  

QAP thanks the Hong Kong Research Grants Council for support under GRF grants 17326116 and 17300417. FL and AR thank HKU for postdoctoral fellowships. AAZ thanks the Hung Hing Ying Foundation for the provision of a visiting professorship at HKU. MAG was funded under grant number PGC2018-102184-B-I00 of the Ministerio de Educaci\'on, Innovaci\'on y Universidades co-funded with FEDER funds. This work was partially funded by Kepler/K2 grant J1944/80NSSC19K0112 \& HST GO-15889, and STFC grants  ST/T000414/1, ST/T000198/1,  ST/S006109/1 and  by the EU HORIZON2020 program under grant 101004214.

Part of this work is based on data from (i) the OMC Archive at CAB (INTA-CSIC) pre-processed by ISDC; (ii) data products from the Near-Earth Object Wide-field Infrared Survey Explorer (NEOWISE) funded by NASA; (iii)  the Swift public data archive; (iv) the NASA/IPAC Infrared Science Archive; ( v)  the AAVSO Photometric All-Sky Survey (APASS); (vi) data from the Asteroid Terrestrial-impact Last Alert System (ATLAS) project; (v) Extinction data from the EXPLORE platform funded by the EU under grant 101004214; (vi) {\it Gaia} spectroscopy from the Gaia Data Mining Platform at the Royal Observatory Edinburgh, funded by STFC. 

For the purpose of open access, the authors have applied a Creative Commons Attribution (CC BY) licence to any Author Accepted Manuscript version arising. This publication is supported by multiple datasets which are openly available at locations cited in the References.


%

\vspace{5mm}
\facilities{GTC (OSIRIS), WIYN (SparsePak), XMM, Swift (XRT), WISE, AKARI, TESS, IRSA, Planck, IRAS, ROSAT, VOSA, HST (STIS), ATLAS, Gaia}


\software{astropy \citep{2013A&A...558A..33A,2018AJ....156..123A}, 
cmfgen \citep{hillier1998,hillier2012a,hillier2012b} (\url{http://kookaburra.phyast.pitt.edu/hillier/web/CMFGEN.htm})
cloudy \citep{2013RMxAA..49..137F}, pyphot (\url{https://github.com/mfouesneau/pyphot} ), GaiaXPy (\url{https://gaia-dpci.github.io/GaiaXPy-website/})
          }

\bibliography{new_Pa30_biblio}{}
\bibliographystyle{aasjournal}

\end{document}